\documentclass[lettersize,journal]{IEEEtran}
\usepackage{amsmath,amsfonts}
\usepackage{algorithm}
\usepackage{algpseudocode}
\usepackage{array}
\usepackage[caption=false,font=normalsize,labelfont=sf,textfont=sf]{subfig}
\usepackage{textcomp}
\usepackage{stfloats}
\usepackage{url}
\usepackage{verbatim}
\usepackage{graphicx}
\usepackage{cite}
\usepackage{tikz}
\usepackage{amsmath,amsfonts}
\usepackage{algorithm}
\usepackage{algpseudocode}
\usepackage{array}
\usepackage[caption=false,font=normalsize,labelfont=sf,textfont=sf]{subfig}
\usepackage{textcomp}
\usepackage{stfloats}
\usepackage{url}
\usepackage{verbatim}
\usepackage{graphicx}
\usepackage{cite}
\usepackage{tikz}
\usetikzlibrary{fit}
\usepackage{setspace}
\usepackage{graphicx}
\usepackage{multirow,multicol}
\usepackage[]{amsmath}
\usepackage{amsbsy}
\usepackage{amssymb}
\usepackage{amsfonts}
\usepackage{pifont}
\usepackage{balance}
\usepackage{fancyvrb}
\usepackage[framemethod=TikZ]{mdframed}

\definecolor{lightgray}{RGB}{220,220,220}
\definecolor{myblue}{cmyk}{1 0.35 0 0.5}
\definecolor{myyellow}{cmyk}{0.5 0.1 0.9 0.1}

\newtheorem{definition}{Definition}

\newenvironment{densitemize}
{\begin{list}               
    {$\bullet$ \hfill}{
        \setlength{\leftmargin}{\parindent}
        \setlength{\parsep}{0.04\baselineskip}
        \setlength{\itemsep}{0.5\parsep}
        \setlength{\labelwidth}{\leftmargin}
        \setlength{\labelsep}{0em}}
    }
{\end{list}}

\newcommand{\boxedthm}[1]{
\begin{mdframed}[roundcorner=1pt,middlelinewidth=1pt,backgroundcolor=lightgray!20]
#1
\end{mdframed}
}

\providecommand{\eref}[1]{Eq. \eqref{#1}}  
\providecommand{\cref}[1]{Chapter~\ref{#1}}

\providecommand{\fref}[1]{Figure~\ref{#1}}
\providecommand{\tref}[1]{Table~\ref{#1}}

\providecommand{\R}{\ensuremath{\mathbb{R}}}

\providecommand{\E}{\ensuremath{\mathbb{E}}}

\renewcommand{\vec}[1]{\ensuremath{\boldsymbol{#1}}}

\providecommand{\calB}{\mathcal{B}}

\providecommand{\calN}{\mathcal{N}}

\providecommand{\calP}{\mathcal{P}}

\providecommand{\vx}{\mathbf{x}}
\providecommand{\vy}{\mathbf{y}}

\providecommand{\valpha}{\vec{\alpha}}

\providecommand{\vtheta}{\vec{\theta}}

\providecommand{\vtau}{\vec{\tau}}

\providecommand{\blue}[1]{\textcolor{blue}{#1}}

\usepackage{lscape}

\usepackage{tikz}
\usetikzlibrary{calc}
\usepackage{enumitem}

\begin{document}

\title{Spatially Varying Exposure with 2-by-2 Multiplexing: 
Optimality and Universality}

\author{Xiangyu~Qu,~\IEEEmembership{Student~Member,~IEEE},
Yiheng~Chi,~\IEEEmembership{Student~Member,~IEEE}, and~Stanley~H.~Chan,~\IEEEmembership{Senior~Member,~IEEE}%
\thanks{The authors are with the School of Electrical and Computer
Engineering, Purdue University, West Lafayette, IN 47907, USA. Email: {
\{qu27, chi14, stanchan\}}@purdue.edu. This work is supported, in part, by the National Science Foundation under grant 2030570.} %
}

\maketitle

\begin{abstract}
The advancement of new digital image sensors has enabled the design of exposure multiplexing schemes where a single image capture can have multiple exposures and conversion gains in an interlaced format, similar to that of a Bayer color filter array. In this paper, we ask the question of how to design such multiplexing schemes for \emph{adaptive} high-dynamic range (HDR) imaging where the multiplexing scheme can be updated according to the scenes. We present two new findings.

(i) We address the problem of \emph{design optimality}. We show that given a multiplex pattern, the conventional optimality criteria based on the input/output-referred signal-to-noise ratio (SNR) of the independently measured pixels can lead to flawed decisions because it cannot encapsulate the location of the saturated pixels. We overcome the issue by proposing a new concept known as the spatially varying exposure risk (SVE-Risk) which is a pseudo-idealistic quantification of the amount of recoverable pixels. We present an efficient enumeration algorithm to select the optimal multiplex patterns.

(ii) We report a \emph{design universality} observation that the design of the multiplex pattern can be decoupled from the image reconstruction algorithm. This is a significant departure from the recent literature that the multiplex pattern should be jointly optimized with the reconstruction algorithm. Our finding suggests that in the context of exposure multiplexing, an end-to-end training may not be necessary.
\end{abstract}


\section{Introduction}
\IEEEPARstart{D}{igital} image sensors today, at least for the majority of them, pick and choose a \emph{global} exposure and conversion gain across the entire pixel array to control the amount of photon flux reaching the sensor. For high dynamic range (HDR) scenes, this global configuration requires the camera to capture a bracket of exposures and use post-processing algorithms to fuse an HDR image. However, in the presence of motion and noise, HDR fusion is known to be difficult.

Approximately two decades ago, Nayar and Mitsunaga proposed the idea of spatially multiplexing the exposure and conversion gain \cite{Nayar_2000_SVE}. The argument was that we could capture multiplexed exposures like color filter arrays in a single-shot to avoid the motion problem. The reduction of the spatial resolution can be, in principle, recovered by an appropriately designed interpolation algorithm. Nayar and Mitsunaga's idea led to a series of very interesting work in coded exposures, including adaptive control schemes \cite{Nayar_2003_adaptive}, hardware multi-bucket sensor designs \cite{Wan_2012_Multi-bucket}, and some of the most recent works in co-optimizing the multiplex pattern and the reconstruction algorithm via deep learning \cite{Nguyen_2022_ICCP,Klinghoffer_2022_review}.

\begin{figure}[t]
  \centering
   \includegraphics[width=1\linewidth]{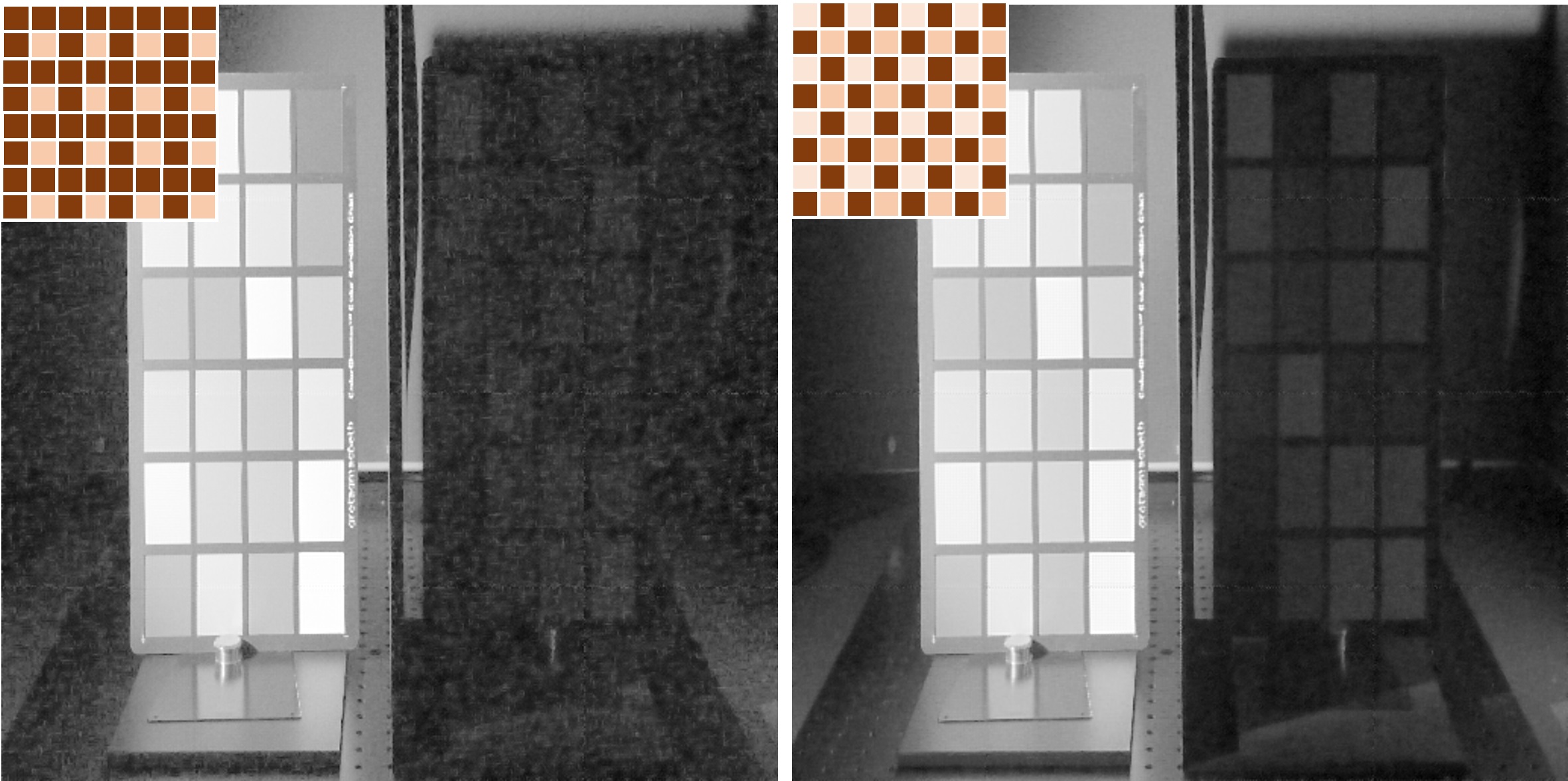}
   \caption{Given an HDR scene, the choice of spatially-varying exposure pattern significantly influences the quality of the reconstructed image. We propose a new risk estimator to determine the optimal exposure pattern for capturing high dynamic range scenes. \textbf{[Left]} HDR scene reconstruction using an arbitrary spatially-varying exposure pattern. \textbf{[Right]} HDR scene reconstruction using the optimal exposure pattern for the scene. }
   \label{fig: Fig01}
\end{figure}

Despite the large number of prior work, we seldom ask the question of how to \emph{design} the multiplex pattern. This is a meaningful question, because a poorly chosen multiplex pattern can produce a substantially worse image than the optimal one, as illustrated in \fref{fig: Fig01}. However, if we want to choose the optimal pattern, we must first answer the question:
\boxedthm{
What is the \emph{optimality criteria} for exposure multiplex?
}
``Optimality criteria'' may seem trivial --- just pick a pattern that maximizes the signal-to-noise ratio (SNR)! But SNR of what? If it is the SNR of the \emph{measured} pixel values, then we need a way to quantify the locations of the saturated pixels: a group of sparsely located saturated pixels are easy to recover (think about the color filter arrays) whereas a group of densely concentrated pixels are hard to recover (think about inpainting a large hole in an image). How about the SNR of the \emph{reconstructed} image? This seems more plausible because if a final reconstruction has the highest PSNR, then the pattern is optimal. However, to compare the final reconstruction of every pattern, we need to first capture the scene using every possible pattern. This exhaustive capturing and reconstruction approach defeats the purpose of finding a good multiplex pattern; if one already has all possible captures and associated reconstructions, why bother finding the pattern to be used? Therefore, as we can see, the above seemingly easy task can quickly evolve into a challenging research problem.

The goal of this paper is to clarify the difficulties and propose solutions. Our two contributions are:

\begin{densitemize}
\item We introduce the concept of spatially varying exposure risk (SVE-Risk). SVE-Risk is customized to measure the usefulness of a multiplex pattern. It is universal in the sense that the risk does not require knowledge about a particular image reconstruction algorithm but can still be used to predict what pattern will likely yield the optimal final image for the given scene. We also propose efficient computation methods for SVE-Risk and demonstrate practicality of using our suite of algorithms for multiplex pattern selection.
\item We discover, through large-scale experiments, that a multiplex pattern and the image reconstruction algorithm do not need to be co-optimized. This is a departure from prior work that argue the necessity of co-optimization. We hope that our finding can stimulate discussions about the future sensor-algorithm co-design problems.
\end{densitemize}

\section{Problem Statement and Related Work}
In this section we define notations, state the problem, and comment on the related work.

\subsection{Problem Statement and Notations}
\label{sec:PS}
We let $\vtheta = [\theta_1,\ldots,\theta_N]\in \R^N$ be the ground truth image denoting the incoming photon flux, and let $\vy \in \R^N$ be the observed photoelectric signal produced by the sensor. If the sensor uses a global exposure $\tau$ and a global gain $\alpha$ to capture the image $\vy$, then $\vy$ can be simulated according to the equation
\begin{align}
\vy(\alpha,\tau)
&= \text{ADC}\left\{\text{Clip}\left\{\text{CRF}\left\{ (\calP[\tau\vtheta \cdot \text{QE}] + \tau\mu_{\text{dark}})  \cdot \alpha \right\}\right\} \right. \notag \\
&\left. \qquad + \calN(0,\sigma_{\text{read}}^2) \right\} .
\label{eq: main sensor model}
\end{align}
We regard \eref{eq: main sensor model} as the first order approximation to the actual image sensor. The symbols we used here are defined in \tref{tab:sensor_model}.

\begin{table}[h]
\caption{Image sensor model parameters}
\resizebox{\columnwidth}{!}{
\begin{tabular}{lll}
\hline
Symbol & Meaning & Typical Value\\
\hline
ADC & Analog-digital  & 10\textasciitilde12-bit \\
Clip& Full well limit & 5000 e- \\
$\tau$ & Exposure     & 10ms, 20ms, 40ms\\
$\mu_{\text{dark}}$ & dark current & 0.002 e-/s\\
QE  & Quantum efficiency & 80\%\\
$\alpha$ & Conversion gain & 1 \\
$\sigma_{\text{read}}$ & Read noise & 0.3 e-\\
CRF & Camera Response Function & \\
$\calP$ & Poisson distribution & \\
$\calN$ & Gaussian distribution & \\
\hline
\end{tabular}
}
\label{tab:sensor_model}
\end{table}

The concept of multiplexing is to assign, periodically, an exposure pattern and a gain pattern such that each pixel will be subject to a different exposure and gain. Mathematically, for a $2 \times 2$ exposure and gain pattern, we define
\begin{align*}
\vtau = \{\tau_{\ell}\} =
\begin{bmatrix}
\tau_1 & \tau_2 \\
\tau_3 & \tau_4
\end{bmatrix}
\qquad
\valpha = \{\alpha_{\ell}\} =
\begin{bmatrix}
\alpha_1 & \alpha_2 \\
\alpha_3 & \alpha_4
\end{bmatrix}
\end{align*}
where each $\tau_\ell$ is sampled from a set of exposure levels, e.g., $\{1,2,4,8,...\}$. The same holds for the gain $\alpha_{\ell}$.

The core research question we ask in this paper is the choice of $\vtau$ and $\valpha$.
\boxedthm{
Given the scene radiance $\vtheta$, how do we select $\vtau = \{\tau_\ell\}$ and $\valpha = \{\alpha_\ell\}$ to generate a $\vy$ such that the reconstructed image $\widehat{\vtheta}(\vy)$ has the highest PSNR?
}

\textbf{Why limit to $2\times 2$ patterns?} We limit the scope of this paper to $2 \times 2$ multiplex patterns. Readers may say: This is too restrictive. Why not analyze $4 \times 4$ or $16 \times 16$? Our short answer is that $2 \times 2$ patterns are more hardware friendly than other options.\footnote{An analogy worth mentioning is the color filter arrays: While we all agree that the $2 \times 2$ Bayer pattern is sub-optimal, today we only see a handful of non-Bayer color filters in camera products. Even the latest 4-cell Quad-Bayer patterns by Sony, Samsung and OmniVision are just variants of Bayer.} However, even so, the purpose of this paper is not to argue that $2 \times 2$ is the best option. Instead, the question we ask is that \emph{given} the problem of using $2 \times 2$, how to determine the optimal one? Our analysis on $2 \times 2$ can be generalized to other pattern sizes.

\textbf{What other options do we have for exposure control}? The theme of this paper is exposure and gains controls. In the literature, there are three mainstream approaches:

\begin{enumerate}
    \item \textbf{One fixed pattern for all}. Mount a static mask with spatially-varying light transmittance on top of the sensor array. This requires minimal/no additional circuit as compared with traditional CMOS sensors. However, the spatially-varying exposure pattern is then fixed and cannot be changed to adapt to the scene. Most traditional work on spatially-varying exposure image reconstruction explicitly/implicitly use this approach \cite{Nayar_2000_SVE, aguer_2014, LPA, michael_2012, cogalan_2022, yasuma_2010, shifted_coded_mask, sve_quad_bayer}.
    \item (This paper) \textbf{Periodic $2 \times 2$, updated on-the-fly}. Use different signal lines to control the transfer timing (and hence control the exposure time) of different pixels. This approach allows the spatially-varying exposure pattern to be set on-the-fly and adapt to the scene. However, the number of signal lines increases linearly with the number of independent controls, and so the number of signal lines needs to be small. Overall, it is more functionally versatile than mounted static mask at the cost of some additional circuitry.
    \item \textbf{Per-pixel or per-block, updated on-the-fly}. Add in-pixel latches/flip-flops/logic circuit (also called digital pixel sensor as compared with traditional active pixel sensor). This setup allows maximal level of flexibility. Usually, the exposure of each pixel can be independently controlled. The downside of this approach is that the required additional circuitry is gigantic and that it results in a low fill factor (as low as $6\%$ \cite{scamp5}) as well as higher circuit noise. Most of the focal plane coded exposure work adopt this approach \cite{luo2018,scamp5,ke2019extending,programmable_pixel_array_patent,zhang_patent_2019,dual_tap_cep,neural_sensors,zhang2020}.
\end{enumerate}
Among the three options, we do not prefer Option A because if we want it to be applicable to all images, then it must optimize for the average case. This will very likely lead to a periodic pattern. Option B is preferred over Option C because its hardware requirement is lower.

\subsection{Related Work}
Our work stands at the intersection of spatially-varying-exposure (SVE) imaging and HDR imaging exposure control. Some existing works are worth noting.

\textbf{SVE Imaging:} After Nayar and Mitsunaga \cite{Nayar_2000_SVE}, a number of methods have been proposed to: improve image reconstruction quality \cite{aguer_2014}, improve reconstruction speed as well as robustness against non-uniform noise strength \cite{kronander_2013, LPA}, adapt the application of SVE HDR imaging from a single image to videos \cite{michael_2012}, leverage the higher single frame dynamic range advantage of SVE imaging to tackle motion registration problem of HDR video capturing \cite{cogalan_2022}, extend the original concept of SVE to the idea of generalized assorted pixel \cite{yasuma_2010}, in which image resolution, dynamic range, and spectral profile can be balanced post-capture by imaging with an optimized complex SVE mask. 

Among these SVE imaging works, \cite{yasuma_2010} is closest to our problem and we emphasize the difference between their problem setting and ours in Table~\ref{tab:context_diff}.

\begin{table}[h]
\footnotesize
\caption{Comparison of problem settings}
\begin{tabular}{l|l|l}
\hline
 & \textbf{Yasuma et al. \cite{yasuma_2010}} & \textbf{Ours} \\
\hline
Pattern & \underline{fixed} & \underline{adjustable} on-the-fly\\
\hline
Optimization & optimal for the \underline{average}  & optimal for every \underline{single} \\
 & over all lighting scenarios &  scene \\
\hline
Objective & obtain a universal pattern & select best for current scene\\
\hline
\end{tabular}
\label{tab:context_diff}
\vspace{1ex}
\end{table}

\textbf{HDR Exposure Control:} Exposure bracketing is a very popular HDR imaging technique, which fuses multiple LDR images exposed at different levels to form one HDR image, for example, imaging with dual sampling sensors \cite{yadid1997wide}, on-chip fusion \cite{wang2001high}, and recently using deep networks \cite{kalantari2017deep, wu2018deep, yan2019attention, yan2019multi, yan2020deep, deng2020multi, yan2022lightweight, zhu2022hdrfeat, chi2023hdr}. Despite its popularity, few studies focus on selecting proper exposure levels for the LDR frames. These exposure control algorithms devised for exposure bracketing mostly fall into two groups: (1) algorithms focusing on design simplicity and efficiency so as to be deployed on imaging devices \cite{pourreza_2015, barakat_2008, guthier_2012, huang_2013}, and (2) image formation modeling based algorithms focusing on optimality, i.e., to find the optimal set of exposures by some metric for reconstructing the scene \cite{hirakawa_2010, hasinoff_2010}. A recent work also demonstrates the possibility of using reinforcement learning to train an exposure bracketing selection network \cite{wang2020learning}. In terms of design, our SVE pattern selection algorithm is an image formation modeling based algorithm; however, methods like \cite{hirakawa_2010, hasinoff_2010} cannot be migrated to our problem without significant changes, in that spatial multiplexing and pixel interpolation are not within their problem scopes. Imaging with other types of image sensors, e.g., Quanta Image Sensors (QIS) \cite{chi2020dynamic, li2021photon, gnanasambandam2020hdr} and Single-Photon Avalanche Diode (SPAD) \cite{ingle2021passive}, are also candidate solutions to HDR imaging. It is further suggested by \cite{chan2022does, chan2023insensitivity} that low bit-depth sensors provide wider dynamic range. In this work, we use a general sensor model \eref{eq: main sensor model}.

\vspace{3ex}
\section{SVE Pattern Selection for HDR}
\label{sec:algorithms}
In this section we present the core idea of this paper, which is the concept of SVE-Risk and efficient methods to evaluate the SVE-risk.

\subsection{Limitations of SNR}
\label{sec:algorithms:snr-limitation}

To motivate the definition of the SVE-Risk, we first discuss the limitations of the per-pixel output-referred\footnote{In this paper we are interested in full-well capacities that are sufficiently large. For pixels with extremely small full-well capacity, e.g., Quanta Image Sensors (QIS), one needs to use the more general formula known as the exposure-referred SNR. We refer to the article by Chan \cite{exposure_referred_snr} for details.} signal-to-noise ratio (SNR). In the context of our problem and image formation model, the SNR at pixel $i$ is
\begin{equation}
\text{SNR}_i
=
\begin{cases}
\frac{(\alpha_i\tau_i)\theta_i}{\sqrt{\alpha_{i}^2(\tau_i\theta_i + \mu_{\text{dark}}) + \sigma_{\text{read}}^2}}, \; &(\alpha_i\tau_i)\theta_i \le V_{\text{max}},\\
0, \; & (\alpha_i \tau_i) \theta_i > V_{\text{max}},
\end{cases}
\label{eq: SNRi}
\end{equation}
where $V_{\text{max}}$ is the maximum voltage allowed by the ADC.

Intuitively, the SNR highlights two aspects of the exposure/gain: (i) If the pixel is not saturated, then the SNR will increase with $\alpha_i$ and $\tau_i$. (ii) If the pixel is saturated (so the signal received by ADC exceeds $V_{\text{max}}$), then the SNR is zero. The two cases are consistent with the classical model in \cite{Nayar_2000_SVE}.

With the per-pixel $\text{SNR}_i$ defined, it is straightforward to define the risk of adopting a particular $(\valpha, \vtau)$ for capturing a given scene using average SNR of all pixels:
\begin{equation}
\text{SNR-Risk}(\valpha,\vtau) = \left(\frac{1}{N}\sum_{i=1}^N \text{SNR}_i\right)^{-1},
\label{eq: SNR-Risk}
\end{equation}
where the reciprocal is used to convert the SNR to a risk.

SNR-Risk has two major drawbacks:
\begin{enumerate}
\itemsep0em
    \item[(i)] SNR-Risk is agnostic to how saturated pixels distribute in the image. Consider the example shown in \fref{fig:snr_limitation 1}. While the two patterns will give exactly the same SNR, only pattern A is recoverable because neighboring pixels are available. Pattern B contains a large region of saturated pixels, which is very difficult to recover.
\begin{figure}[h!]
\centering
\vspace{1ex}
\begin{tabular}{c}
\includegraphics[width=0.85\linewidth]{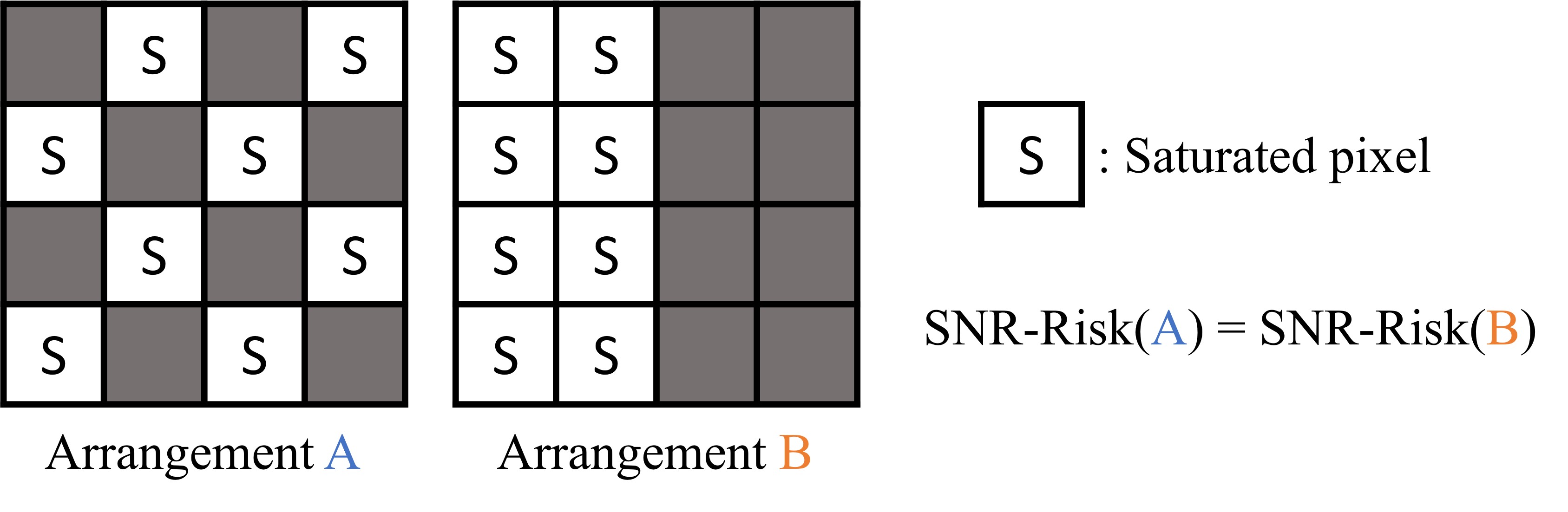}
\end{tabular}
\caption{Two arrangements of saturated pixels share the same SNR-Risk; however, the image captured with arrangement A can be recovered by interpolation while it is more challenging to recover that with arrangement B. }
\label{fig:snr_limitation 1}
\vspace{1ex}
\end{figure}
    \item[(ii)] SNR is a pixelwise calculation without considering its neighbors. Therefore, if there is a bright region, all four control elements of the $2 \times 2$ pattern will try to match the scene without coordinating among themselves. The consequence is that SNR-Risk will choose an all bright or all dark pattern.
\end{enumerate}

The SNR-Risk computes the risk for a pixel depending solely on one pixel. This is in contrast to the fact that practical image reconstruction algorithms almost always aggregate spatial information before predicting a pixel's output value. Hence, SNR-Risk tends to over-estimate the risks associated with easily recoverable pixels. To overcome this issue, we want to design a risk with the neighborhood relations of pixels taken into consideration. To summarize, SNR-Risk is not a good metric because of the following.
\boxedthm{
SNR-Risk cannot comprehend the local structure of the exposure, hence it cannot tell us whether the acquired image is \textbf{recoverable}.
}

\subsection{SVE-Risk}
\label{sec:algorithms:sve-risk:def}
After explaining why SNR is not a good metric to assess the multiplex patterns, in this subsection we introduce a new concept called spatially varying exposure risk (SVE-Risk).

We first consider the definition of an ideal risk. Let $\vy$ be the sensor readout and estimator $\widehat{\vtheta} = [\widehat{\theta}_1,\ldots,\widehat{\theta}_N]$ be the reconstruction mapping that produces an estimate $\widehat{\vtheta}(\vy,\valpha,\vtau)$ of $\vtheta = [\theta_1,\ldots,\theta_N]$. The ideal risk is
\begin{equation}
\small
\text{Risk}(\valpha,\vtau)
= \underset{\widehat{\vtheta}}{\text{inf}} \;\; \E_{\vy}\left[ \frac{1}{N}\sum_{i=1}^{N} \left (\frac{\widehat{\theta}_{i}(\vy, \valpha,\vtau) - \theta_i}{\theta_i} \right )^2 \right],
\label{eq: risk}
\end{equation}
where $\theta_i$ and $\widehat{\theta}_i(\cdot)$ denote the $i$-th element of the ground truth radiance $\vtheta$ and the estimate $\widehat{\vtheta}(\cdot)$, respectively. Note that if the estimator $\widehat{\vtheta}$ is a pixel-wise maximum likelihood (ML) estimator using the forward model defined in \eref{eq: main sensor model}, then the squared ratio in \eref{eq: risk} is exactly the inverse of pixel-wise output-referred SNR for non-overflow pixels. From this perspective, the SNR-Risk can be deemed as a special case of the ideal risk, in which the estimator is predetermined to be a pixel-wise ML estimator.

The caveat of \eref{eq: risk} is that an oracle estimator $\widehat{\vtheta}$ for a scene is never known and it is also impossible to obtain the infimum by enumerating all possible reconstruction algorithms. To mitigate this issue, we approximate the risk by using a hypothetical ideal \emph{local} estimator. This hypothetical estimator cannot be constructed in practice (any reconstruction algorithm is likely to be worse than the hypothetical estimator), but it can give us a meaningful approximation to the infimum.

\begin{definition}[Local estimator]
A \textbf{local estimator} $\widehat{\theta}_i$ at the pixel $i$ is a function that maps the neighborhood observations $\vy_i = \{y_j \,|\, j \in \calB_i \}$ to an estimate $\widehat{\theta}_i(\vy_i)$, where $\calB_i$ denotes the neighborhood around pixel $i$. If $y_i$ is saturated, $\widehat{\theta}_i$ uses the neighborhood \emph{without} $y_i$, i.e., $\vy_{-i} = \{y_j \,|\, j \in \calB_i \}\backslash \{y_i\}$.
\end{definition}

As we define this hypothetical local estimator, we assume that it has the perfect knowledge about inter-pixel correlations of its neighborhood. Therefore, it allows us to achieve two things:
\begin{densitemize}
\item When a pixel is not saturated, the estimator will return us the same value as the SNR-Risk.
\item When a pixel is saturated, the estimator will make an interpolation. The interpolated pixel will have a risk no higher than the largest risk within the neighborhood.
\end{densitemize}

Based on these properties, we can define the SVE-Risk by considering three situations:
\boxedthm{
The \textbf{SVE-Risk of the $i$-th pixel} is defined as
\begin{align}
&\text{SVE-Risk}_{i}(\valpha,\vtau) \label{eq: SVE risk i}\\
&=
\begin{cases}
\frac{1}{\theta_i^2 \lvert \calB_i^*\rvert} \left[\frac{(\theta_i + \mu_{\text{dark}})}{\tau_i} + \frac{\sigma_{\text{read}}^2}{\alpha_i^2\tau_i^2}\right],
& \text{unsaturated},\\
\max\limits_{j \in \calB_i \backslash \{i\}} \frac{1}{\theta_j^2 \lvert\calB_j^*\rvert} \left[\frac{(\theta_j + \mu_{\text{dark}})}{\tau_j} + \frac{\sigma_{\text{read}}^2}{\alpha_j^2\tau_j^2}\right],
& \text{sat., $\checkmark \, \calB_i$},\\
 \left(\frac{V_{\text{max}}}{\alpha_i \tau_i}-\theta_i\right)^2,
& \text{sat., $\times \, \calB_i$},
\end{cases}
\notag
\end{align}
where $\calB_i^* = \calB_i  \setminus \mathcal{S}$ is the neighborhood around pixel $i$ minus the set of saturated pixels $\mathcal{S}$.}

Let's elaborate on the three cases in the definition:
\begin{densitemize}
\item Unsaturated: If a pixel is unsaturated, the risk is defined as the variance of the measurement (which is the denominator of \eref{eq: SNRi}, squared and normalized). The normalization is necessary for preventing bright regions in the image from dominating darker region risks and is crucial for HDR imaging. For unsaturated pixels, the risk calculated in \eref{eq: SVE risk i} is essentially SNR-Risk value scaled by the number of observations in the neighborhood.
\item $\text{Saturated, $\checkmark \, \calB_i$}$: The neighborhood contains pixels that can be used for interpolation. In this case, the risk is defined as the worst variance in the neighborhood. The intuition is that since SVE-Risk uses neighboring pixels to determine the risk, it can be deemed as an extension of the SNR-Risk where we combine independent exposures.
\item \text{Saturated, $\times \, \calB_i$}: The neighborhood does not contain any useful pixels. The risk is defined as the squared error between the cutoff $\frac{V_{\text{max}}}{\alpha_i\tau_i}$ and expected radiance $\theta_i$. Note that this risk is worse than the second case because, in the second case, the substitution comes from one of the unsaturated pixels. For the third case, the difference between $\frac{V_{\text{max}}}{\alpha_i\tau_i}$ and $\theta_i$ can be very large if $\theta_i$ is far from the cutoff.
\end{densitemize}

With pixel-wise SVE-Risk defined, we define overall SVE-Risk by taking sum of each pixel's risk.
\boxedthm{
The \textbf{SVE-Risk of the whole image} is defined as
\begin{align}
\text{SVE-Risk}(\valpha,\vtau) = \sum_{i=1}^N \text{SVE-Risk}_{i}(\valpha,\vtau).
\label{eq:sve_risk_total}
\end{align}
}

\subsection{Computing SVE-Risk}
\label{sec:algorithms:sve-risk:computation}
\begin{figure}[h]
\centering
\includegraphics[width=\linewidth]{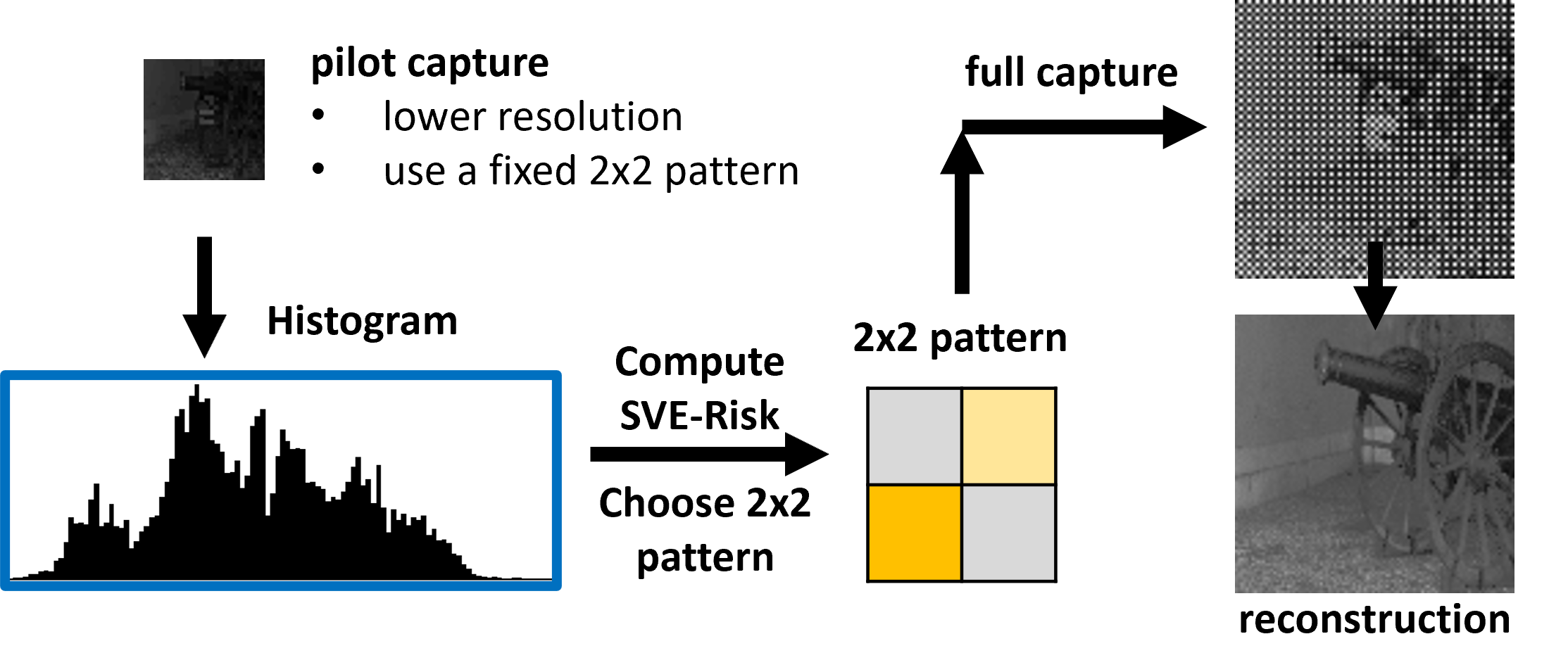}
\caption{Our imaging pipeline is for \emph{dynamical} control of the exposure. We capture a low-resolution pilot image using a fixed $2 \times 2$ exposure pattern (typically high-medium-medium-low) to construct a histogram. From the histogram we compute the risk, and guide by the risk we capture the full image. Since the SVE-Risk is calculated based on the histogram instead of the image, its computational cost is very low. }
\label{fig: pipeline}
\end{figure}

After defining the SVE-Risk, the next big question is how to compute this seemingly ``uncivilized'' \eref{eq: SVE risk i}. However, before we explain how we calculate the SVE-Risk, we first explain the overall imaging pipeline outlined in \fref{fig: pipeline}.

Since our imaging goal is to dynamically control the exposure, our decision on the $2\times 2$ pattern needs to be fast. As illustrated in \fref{fig: pipeline}, the way we compute the SVE-Risk is based on the histogram of the scene radiance. This histogram does not need to be perfectly precise. Therefore, we can use a lower resolution of the same scene, and we can use any average $2 \times 2$ pattern such as high-medium-medium-low. The purpose of the pilot capture is to construct a histogram so that we can make decisions for choosing the $2 \times 2$ pattern. Once we have the $2\times 2$ pattern determined, we perform a full capture and image reconstructions.

\begin{figure}[!h]
\centering
\includegraphics[width=\linewidth]{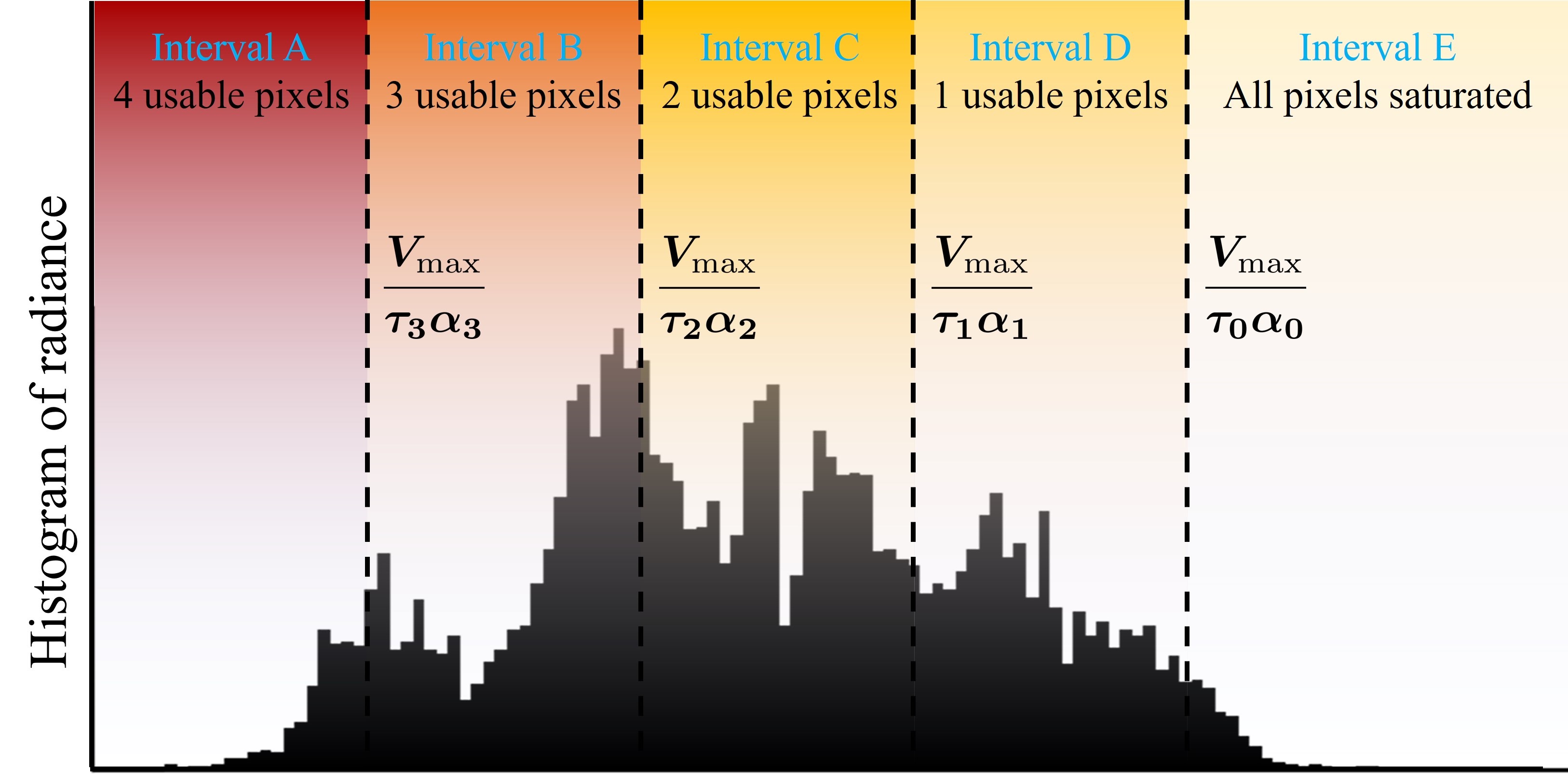}
\caption{An example radiance histogram. The overall radiance histogram is partitioned by the 4 exposures/gains into 5 intervals. For the unsaturated pixels in intervals A to D, the risk associated with them corresponds to the unsaturated case in \eref{eq: SVE risk i}. Pixels in the interval above the threshold $V_\text{max} / \tau_0\alpha_0$ (interval E) are all saturated and unlikely to be recovered. Their risks are computed as the empirical mean squared error as the saturated, $\times \, \calB_i$ case in \eref{eq: SVE risk i}. Risks of saturated pixels in intervals below $V_\text{max} / \tau_0\alpha_0$ (intervals A to D) can be substituted with the worst risks of their neighbors, which is the saturated, $\checkmark \, \calB_i$ case in \eref{eq: SVE risk i}. 
}
\label{fig:sve-risk:computation-visual}
\end{figure}

We now discuss the histogram. Assume we have collected a pilot image. We denote the distribution of the radiance as $p(\theta)$. An example is shown in \fref{fig:sve-risk:computation-visual}. We stress again that operating on the radiance distribution rather than individual pixels is more efficient, because, once the histogram is built, the complexity of pattern risk estimation becomes proportional to the number of bins in the histogram (order of magnitude is hundreds to thousands) instead of the number of pixels (order of magnitude is hundreds of thousands to millions).

Given a pattern $\{(\tau_\ell, \alpha_\ell) \,|\, \ell = 0,\ldots,3\}$, without loss of generality, we assume that the exposure/gains are sorted so that $\alpha_0 \tau_0 \le \alpha_1 \tau_1 \le \alpha_2 \tau_2 \le \alpha_3 \tau_3$. The pattern gives us four cutoff radiance levels $\frac{V_{\text{max}}}{\alpha_\ell \tau_\ell}$ which partition the entire radiance range into five intervals, as shown in \fref{fig:sve-risk:computation-visual}.

As shown in \fref{fig:sve-risk:computation-visual}, a radiance level $\theta$ saturates an exposure/gain element $(\tau_l, \alpha_l)$ if it is greater than the corresponding cutoff $\frac{V_{\text{max}}}{\alpha_\ell \tau_\ell}$. For $\theta > \frac{V_{\text{max}}}{\alpha_0 \tau_0}$ (interval E), they saturate all exposure/gain elements and it is very likely that the pixels corresponding to these radiance levels are not recoverable by any reconstruction algorithm. In this case, risk associated with these radiance levels is, according to Case 3 in \eref{eq: SVE risk i}:
\begin{align}
\overline{\text{SVE-Risk}}_{\text{nonrecoverable}}(\valpha,\vtau, \theta) = \left(\frac{V_{\text{max}}}{\alpha_0 \tau_0}-\theta\right)^2 .
\label{eq: SVE risk 3.3a}
\end{align}

Now consider a radiance level between two cutoffs $\frac{V_{\text{max}}}{\alpha_{\ell} \tau_{\ell}} < \theta < \frac{V_{\text{max}}}{\alpha_{\ell-1} \tau_{\ell-1}}$ (i.e., a radiance level in intervals B, C, D). The radiance level $\theta$ saturates all exposure/gain elements below it; however, since there are neighboring elements not saturated by $\theta$, it is likely that saturated pixels at this radiance level can be recovered by exploring their neighbor pixels. The risk associated with this radiance level is then, according to Case 1 and Case 2 in \eref{eq: SVE risk i}:
\begin{align}
&\overline{\text{SVE-Risk}}_{\text{recoverable}}(\valpha,\vtau, \theta) \notag\\
&=
\frac{1}{4} \left[ \underset{ \text{\textcolor{blue}{contribution from non-saturated elements}} }{ \underbrace{\sum_{j=0}^{\ell-1} \frac{1}{B_j\theta^2} \, \left(\frac{(\theta + \mu_{\text{dark}})}{\tau_{j}} + \frac{\sigma_{\text{read}}^2}{\alpha_{j}^2\tau_{j}^2}\right)} } \notag \right. \\
&\left. \qquad + \underset{ \text{\textcolor{blue}{saturated, use risk of worst non-saturated neighbor}} }{ \underbrace{\sum_{j=\ell}^{3} \frac{1}{B_0\theta^2} \, \left(\frac{(\theta + \mu_{\text{dark}})}{\tau_{0}} + \frac{\sigma_{\text{read}}^2}{\alpha_{0}^2\tau_{0}^2}\right)} }
 \right] ,
\label{eq: SVE risk 3.3}
\end{align}
where $B_j$ is the number of non-saturated pixels within the neighborhood (with predetermined size) of exposure/gain element $(\alpha_l, \tau_l)$. Note that, since we know how pattern is tiled across the entire sensor array, $B_l$ can be calculated for each saturation scenario (i.e., 0, \dots, 3 exposure/gain elements saturated) beforehand and be stored in memory (see supplementary).

The risk is therefore the sum of \eref{eq: SVE risk 3.3a} and \eref{eq: SVE risk 3.3}.
\boxedthm{
To \textbf{numerically compute the SVE-Risk}, we construct the radiance histogram (like \fref{fig:sve-risk:computation-visual}), and calculate
\begin{align*}
\small
&\text{SVE-Risk}(\valpha,\vtau)\\ &= \int_{\frac{V_{\text{max}}}{\alpha_0 \tau_0}}^{\infty} \, \overline{\text{SVE-Risk}}_{\text{nonrecoverable}}(\valpha, \vtau, \theta) p(\theta) \, d\theta\\
&\qquad + \int_{0}^{\frac{V_{\text{max}}}{\alpha_0 \tau_0}} \overline{\text{SVE-Risk}}_{\text{recoverable}}(\valpha,\vtau, \theta) p(\theta)\, d\theta,
\end{align*}
where $p(\theta)$ is the radiance histogram of the pilot capture.
}

We show a comparison of the run time of calculating the SVE-Risk and the SNR-Risk in Table \ref{tab:running_time}. Since SVE-Risk is calculated using the histogram, it is insensitive to the image resolution. In contrast, since SNR needs to evaluate every single pixel, its computation grows with the number of pixels.

\begin{table}[h]
\centering
\footnotesize
\caption{Comparison of risk estimator run time for sensors of different resolutions. Numbers shown are run time of evaluating the corresponding risk of 495 patterns averaged over 100 executions.}
\begin{tabular}{ccc}
\hline
\hline
Resolution & Runtime for          & Runtime for \\
           & calculating SVE-Risk & calculating SNR-Risk \\
 \hline
512x896   & \textbf{0.20$\pm$0.01 sec} & 0.69$\pm$0.01 sec \\
1024x1792 & \textbf{0.27$\pm$0.01 sec} & 6.13$\pm$0.22 sec\\
2048x3584 & \textbf{0.52$\pm$0.01 sec} & 27.35$\pm$0.12 sec\\
\hline
\end{tabular}
\label{tab:running_time}
\end{table}

\subsection{Efficient Pattern Enumeration}
\label{sec:algorithms:pattern_enum}
There is one final design question we need to answer before using the SVE-Risk. It is the problem of candidate patterns to evaluate. Suppose we have 9 exposure levels for a $2 \times 2$ pattern, we will have a total of $9^4 = 6561$ candidates. It will be too much computation if we need to calculate the SVE-Risk for each candidate pattern. Therefore, in this subsection, we present a method to eliminate low priority patterns.

To remove the low-priority patterns, we made an observation that the majority of all candidate patterns are redundant. For example, from the reconstruction point-of-view, an exposure pattern $\vtau = \{1, 10, 10, 10\}$ is almost identical to patterns $\{10,1,10,10\}$, $\{10, 10, 1, 10\}$, and $\{10, 10, 10, 1\}$. To justify this claim, we experiment with three representative reconstruction algorithms on sensor readouts synthesized with these four patterns. We test on a dataset containing 46 images and show the results in \fref{fig:redundant_pat_example}. Across the four exposure patterns, the reconstruction results are identical for any fixed algorithm.

\begin{table}[!h]
\setlength\tabcolsep{3pt}
\renewcommand{\arraystretch}{1}
\centering
\caption{$\mu$PSNR (dB) of the reconstructed images for the four equivalent multiplex patterns. The PSNR is averaged across 46 test images and rounded to first decimal place. }
\begin{tabular}{ccccc}
\hline
\hline\\
[-0.5em]
&
\includegraphics[width=0.15\linewidth]{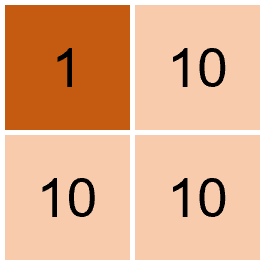}&
\includegraphics[width=0.15\linewidth]{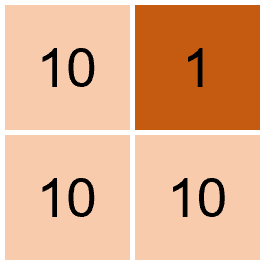}&
\includegraphics[width=0.15\linewidth]{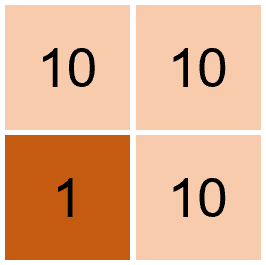}&
\includegraphics[width=0.15\linewidth]{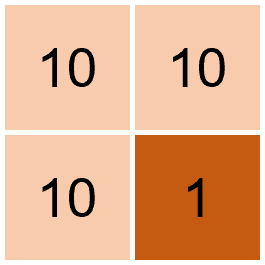}\\
\hline
ADMM-TV\cite{pnp_admm}   & 31.1dB & 31.1dB & 31.1dB & 31.1dB\\
Restormer\cite{Zamir2021Restormer} & 30.0dB & 30.0dB & 30.0dB & 30.0dB\\
LPA\cite{LPA}   & 29.7dB & 29.7dB & 29.7dB & 29.7dB\\
\hline
\end{tabular}
\label{fig:redundant_pat_example}
\end{table}

Based on the observations above, we define the concept of pattern equivalence: \textit{two patterns are equivalent for capturing a scene if they are permutations of each other}. Our proposed strategy is to enumerate on \textbf{pattern equivalence classes} and compute canonical form of each class. The canonical pattern is the one with maximum variation in the $2 \times 2$ grid. Specifically, if the input pattern is $\vtau = [\tau_1,\tau_2,\tau_3,\tau_4]$, we sort the sequence to obtain $\texttt{sort}(\vtau) = \{\tau_{[1]}, \tau_{[2]}, \tau_{[3]}, \tau_{[4]}\}$ where $\tau_{[1]} \le \tau_{[2]} \le \tau_{[3]} \le \tau_{[4]}$. The canonical pattern is then defined as
\begin{equation*}
\texttt{canon}(\vtau) =
\begin{bmatrix}
\tau_{[1]} & \tau_{[3]} \\
\tau_{[4]} & \tau_{[2]}
\end{bmatrix}.
\end{equation*}
The intuition here is that since $\texttt{sort}(\vtau)$ is already sorted, the alternating allocation of the exposure values will maximize the variation within the $2 \times 2$ grid. For example, the canonical forms of the patterns $\vtau = [8, 32, 1, 10]$ and $\vtau = [1, 10, 1, 10]$ are
\begin{equation*}
 \texttt{canon}(\vtau) = \begin{bmatrix}
 1 & 10 \\
 32 & 8
 \end{bmatrix}
 \quad\mbox{and}\quad
 \texttt{canon}(\vtau) = \begin{bmatrix}
 1 & 10 \\
 10 & 1
\end{bmatrix},
\end{equation*}
respectively. Once all the patterns are converted to their canonical forms, checking the equivalence is simplified to check whether the two canonical forms are identical.

By enumerating on equivalence class instead of all realizable patterns, we reduce the complexity to $\sum_{k=1}^{\text{min}(L, m)} \binom{L}{k} \binom{m-1}{k-1}$ while maintain same coverage of realizable pattern space\footnote{See supplementary for the derivation and pseudo-code.}. When $L = 9$ and $m = 4$, this reduces enumeration size from 6561 to 495.

\textbf{Remark}: Readers may ask: The proposed algorithm is largely an exhaustive search and it requires knowledge about the radiance distribution. Is it possible to improve the search? We note that the pilot estimate is designed to be \emph{coarse}. As long as the shape of the radiance distribution is obtained, we can perform the histogram-based calculation. For faster algorithms, we do not think the typical gradient-based algorithm would work here because our problem is discrete with many stationary points. There might be some discrete optimization methods. We are open to explore them in our future work.

\section{Optimality and Universality}
\label{sec:experiments}

In the beginning of the paper, we mentioned two key findings of this paper. Firstly, we claim that for exposure multiplexing, there exists a better \emph{optimality} criteria than the SNR. We have elaborated on the SVE-Risk in the previous section. In this section, we evaluate SVE-Risk by justifying the following statement.
\boxedthm{
\textbf{Optimality}. The SVE-Risk optimal exposure multiplex pattern can generate a raw image, if passed through a reconstruction algorithm, with nearly the highest PSNR.
}
The second claim we made is \emph{universality}. In this section, we justify the following statement.
\boxedthm{
\textbf{Universality}. The optimal exposure multiplex pattern is universally good for \emph{all} image reconstruction algorithms.
}
Because of the empirical nature of both statements, we answer them through experiments. Our experiments involve large-scale datasets, several new ways of visualizing the results, and a collection of real data.

\subsection{Datasets}
\label{sec:experiments:datasets}
Before diving into experiment design and results, we describe datasets and synthesis parameters in more details.
We use linearized ground truth 16-bit HDR images from NTIRE\cite{NTIRE}, HDR-Eye\cite{hdr-eye}, and SIGGRAPH17\cite{siggraph17} datasets as scene radiance maps. These radiance maps are normalized such that, under minimal exposure and unit gain, the photon flux corresponding to 99 percentile is within the ADC range. This setting is practical and realizable on hardware using modern auto-exposure control. We do not attempt to accommodate for the brightest 1\% of pixels, because these pixels usually are light sources that directly shine on the sensor. Each normalized radiance map is resized such that the short edge has a length of 512 pixels. The radiance maps are then used to synthesize raw sensor readout using image formation model in Section \ref{sec:PS}. We use HDR-Eye data for reconstruction algorithms hyper-parameter tuning and model training, SIGGRAPH17 data for hyper-parameter validation, and NTIRE data for testing and empirical study.

\subsection{Reconstruction Algorithms}
\label{sec:experiments:empirical:reconstruction}
An important task of our verification is to evaluate the quality of the \emph{reconstructed} image. Thus, it is necessary to consider image reconstruction algorithms. However, given the sheer volume of reconstruction methods, it would be impossible to evaluate everyone. A surrogate we take here is to consider three representative classes of algorithms. Within each class, we consider a representative method which we can either implement or we have access to the original source code.

\begin{enumerate}
\setlength\itemsep{0ex}
    \item[(i)] \textbf{Non-data-driven non-iterative}: Algorithms in this class do not require any training and usually make relatively simple or even no assumptions about image structure. During reconstruction, the estimation for each pixel is only carried out once (hence non-iterative). Traditional bi-linear/quadratic/cubic interpolation, median filters, filter banks and more fall into this class. We adopt and implement a local polynomial approximation (LPA) \cite{LPA} as a representative of this class.
    \item[(ii)] \textbf{Non-data-driven iterative}: These are classical tools for solving an inverse problem. Compared with non-iterative approaches, algorithms in this class usually model both the forward imaging process and underlying image structure/prior. This class of algorithms alternate between a forward step to handle the data fidelity, and a backward step that integrates the scene prior. Typical examples include Plug-and-Play \cite{pnp_admm} and Piecewise Linear Estimators \cite{yu_2012}, etc. We implement a Plug-and-Play ADMM with total variation prior (ADMM-TV) to represent this category.
    \item[(iii)] \textbf{Data-driven:} Dictionary learning \cite{bao_2013} and neural networks \cite{zhang2017beyond, Zamir2021Restormer, Zamir2021MPRNet} based image reconstruction/restoration methods fall into this category. It should be noted that although some non-data-driven iterative approaches may also adopt a dictionary or network as a sub-component (e.g., one may use a denoiser network in Plug-and-Play framework as prior step), and there has been numerous efforts \cite{deep_unroll, deep_equilibrium} to try to bring together the best of both tools, we do not consider them as purely data-driven approaches. We limit the scope of this category to methods that are one-pass (i.e., non-iterative) and trained directly for the inference task. We use Restormer\footnote{We discovered in our experiment that networks cannot be trained well when the input to a network has a very high dynamic range, and this training failure cannot be saved by input normalization. Therefore, instead of operating on raw sensor readout in linear scale and predict a linear/log scale output (as most of network-based HDR works do. Their inputs are usually LDR images and their task is to combine LDR images into HDR images, so the domain of their problems does not align exactly with ours), we take a log scale normalized sensor readout as input and predict a log scale radiance map.} \cite{Zamir2021Restormer} as an example of this category.
\end{enumerate}

\subsection{Metrics}
Because of the unique problem setting we have, there is no prior standardized evaluation criterion. To this end, we consider a few known metrics and introduce a few new ones.

(i) \textbf{$\mu$PSNR, $\mu$SSIM, $\mu$LPIPS}. In high dynamic range images, high exposure regions can easily dominate losses or metrics over low exposure regions; therefore, evaluating reconstruction quality in linear scale is usually less meaningful. Similar to other HDR related works \cite{wang2020learning}, we evaluate reconstruction quality on $\mu$-tone-mapped images, which is defined as $$\vx_{\mu} = \frac{\log(1 + \vx\cdot\mu)}{\log(1 + \mu)},$$ where $x$ is a linear scale image normalized to $[0, 1]$ and $\mu$ is a hand-picked hyper-parameter controlling the strength of dynamic range compression. In our experiment, we set $\mu$ to the maximum reference level of ADC (see sec. \ref{sec:PS}).

In tone-mapped space, we measure the PSNR ($\mu$PSNR, higher is better), structural similarity ($\mu$SSIM, higher is better) \cite{SSIM}, and perceptual distance ($\mu$LPIPS, lower is better) \cite{zhang2018perceptual} between reconstructions and ground truth images.

(ii) \textbf{SNR-Risk, SVE-Risk, and their variants}. Since one main objective of this paper is to propose SVE-Risk, it is necessary to compare it with SNR. In addition to the standard SNR-Risk and SVE-Risk described in previous sections, we also evaluate following two variants of the risks, which arise naturally as one contemplate why one risk works while the other does not. Thus, we have four risk terms to consider:
\begin{itemize}
    \item SNR-Risk, as defined in \eref{eq: SNR-Risk}.
    \item SVE-Risk, as defined in \eref{eq:sve_risk_total}.
    \item $\text{SNR}_{\text{MSE}}$: SVE-Risk assigns coarse estimates of mean squared error (MSE) as risk to unrecoverable pixels. Given the close relationship between MSE and PSNR/SSIM, one may wonder if this assignment gives SVE-Risk an unfair advantage over SNR-Risk, as SNR-Risk is purely forward model based. To answer this, we modify SNR-Risk by assigning the actual MSE between the normalized sensor readout (i.e., $y/(\alpha \tau)$) and the corresponding ground truth to overflowing pixels. We denote this variant as $\text{SNR}_{\text{MSE}}$ in results.
    \item $\text{SVE}_{\text{w/o}}$: The idea of our SVE-Risk design is that it penalizes neighborhoods with too many saturated pixels through an auto-tuned parameter $|\calB|$. We evaluate SVE-Risk without the penalty term $|\calB|$, denoted $\text{SVE}_{\text{w/o}}$ in results, to show that this idea is indeed imperative for achieving optimal performance instead of being a dubious add-on.
\end{itemize}

\subsection{Verify the Optimality of SVE-Risk}
In this subsection we discuss our experiments to assess the optimality of the SVE-Risk. We first discuss the protocol of the experiment, and then the results.

\textbf{Protocol of experiment}. We would like to compare SNR-Risk and SVE-Risk. The evaluation of SNR-Risk requires access to the radiance map. For convenience, we directly use ground truth radiance map as the input. The evaluation of SVE-Risk is easier because we only need the histogram. We assume we have access to 4 exposures evenly distributed across the total exposure levels. The histogram is then built using all non-saturating pixels.

\begin{table*}[!t]
\caption{Left eight columns: Average reconstruction quality drop (measured by $\mu$PSNR, $\mu$SSIM, $\mu$LPIPS) between using the oracle pattern and the top-1/top-5 pattern(s) for various reconstruction algorithms. Right eight columns: Probability of the reconstruction quality difference between the oracle pattern and the top-1 pattern selected by a risk being greater than pre-determined thresholds. For all categories, smaller numbers correspond to better selections of exposure patterns. The best score in each comparison category is colored in \blue{blue}. We use different Q-score threshold for $\mu$LPIPS because its scale and spread are much wider than those of the other two metrics.}
\resizebox{\textwidth}{!}{%
\begin{tabular}{c|cccc|cccc|cccc|cccc}
\hline
\hline
 & \multicolumn{16}{c}{$\mathbf{\mu PSNR}$}\\
 \cline{2-17}
 & \multicolumn{4}{c|}{$\Delta_1$} & \multicolumn{4}{c|}{$\Delta_5$} & \multicolumn{4}{c|}{$Q(1\%)$} & \multicolumn{4}{c}{$Q(5\%)$}\\
 \cline{2-17}
 & SVE & $\text{SVE}_{w/o}$ & SNR & $\text{SNR}_{mse}$ & SVE & $\text{SVE}_{w/o}$ & SNR & $\text{SNR}_{mse}$ & SVE & $\text{SVE}_{w/o}$ & SNR & $\text{SNR}_{mse}$ & SVE & $\text{SVE}_{w/o}$ & SNR & $\text{SNR}_{mse}$\\
\cline{2-17}

\hline
LPA & \blue{0.66} & 1.03 & 10.32 & 9.66 & 1.04 & \blue{0.98} & 8.5 & 8.81 & \blue{53.8} & 71.4 & 99.9 & 98.4 & \blue{10} & 18.6 & 99.9 & 96.5\\
ADMM-TV & \blue{1.3} & 1.7 & 8.43 & 7.99 & \blue{1.43} & 1.8 & 7.44 & 7.38 & \blue{78.9} & 81.9 & 99.9 & 98.4 & \blue{26.5} & 34.5 & 99.5 & 92.5\\
Restormer & \blue{0.49} & 1.54 & 6.19 & 5.38 & \blue{0.62} & 1.31 & 4.57 & 4.49 & \blue{49.3} & 87.4 & 99.9 & 98.7 & \blue{4.3} & 29.8 & 97.7 & 96\\
\hline

& \multicolumn{16}{c}{$\mathbf{\mu SSIM}$}\\
 \cline{2-17}
 & \multicolumn{4}{c|}{$\Delta_1$} & \multicolumn{4}{c|}{$\Delta_5$} & \multicolumn{4}{c|}{$Q(1\%)$} & \multicolumn{4}{c}{$Q(5\%)$}\\
 \cline{2-17}
 & SVE & $\text{SVE}_{w/o}$ & SNR & $\text{SNR}_{mse}$ & SVE & $\text{SVE}_{w/o}$ & SNR & $\text{SNR}_{mse}$ & SVE & $\text{SVE}_{w/o}$ & SNR & $\text{SNR}_{mse}$ & SVE & $\text{SVE}_{w/o}$ & SNR & $\text{SNR}_{mse}$\\
\cline{2-17}

\hline
LPA & \blue{0.0048} & 0.0050 & 0.1056 & 0.1255 & 0.0093 & \blue{0.0065} & 0.0887 & 0.1096 & \blue{14.3} & 17.4 & 95.9 & 93.6 & 0.3 & \blue{0.1} & 62.2 & 81.3\\
ADMM-TV & 0.0080 & \blue{0.0078} & 0.1099 & 0.1070 & 0.0103 & \blue{0.0098} & 0.0829 & 0.0892 & \blue{27.2} & 29 & 84 & 85.5 & 1.9 & \blue{0.7} & 45.9 & 58.5\\
Restormer & \blue{0.0063} & 0.0106 & 0.0469 & 0.0573 & \blue{0.0075} & 0.0103 & 0.0391 & 0.0443 & \blue{21.6} & 37.4 & 69.5 & 92.2 & \blue{0} & \blue{0} & 22.5 & 54.9\\
\hline

& \multicolumn{16}{c}{$\mathbf{\mu LPIPS}$}\\
 \cline{2-17}
 & \multicolumn{4}{c|}{$\Delta_1$} & \multicolumn{4}{c|}{$\Delta_5$} & \multicolumn{4}{c|}{$Q(20\%)$} & \multicolumn{4}{c}{$Q(80\%)$}\\
 \cline{2-17}
 & SVE & $\text{SVE}_{w/o}$ & SNR & $\text{SNR}_{mse}$ & SVE & $\text{SVE}_{w/o}$ & SNR & $\text{SNR}_{mse}$ & SVE & $\text{SVE}_{w/o}$ & SNR & $\text{SNR}_{mse}$ & SVE & $\text{SVE}_{w/o}$ & SNR & $\text{SNR}_{mse}$\\
\cline{2-17}

\hline
LPA & \blue{0.007} & 0.018 & 0.179 & 0.218 & \blue{0.019} & 0.020 & 0.147 & 0.191 & \blue{10.4} & 40.1 & 99.6 & 97.1 & \blue{0.1} & 2.3 & 83.5 & 94.6\\
ADMM-TV & \blue{0.026} & 0.031 & 0.169 & 0.169 & \blue{0.032} & 0.035 & 0.128 & 0.140 & \blue{58.4} & 71.6 & 87.9 & 97.5 & \blue{8.6} & 13.5 & 72.5 & 78\\
Restormer & \blue{0.03} & 0.046 & 0.08 & 0.112 & \blue{0.035} & 0.044 & 0.070 & 0.093 & \blue{75.8} & 92.6 & 92.2 & 94.5 & \blue{6.4} & 26.2 & 42.6 & 60.5\\
\hline

\hline\hline
\end{tabular}%
}
\label{tab:risk-eval-results}
\end{table*}

Given a set of candidate patterns (495 patterns in our experiment), we define the \textbf{oracle pattern} as the one that gives the highest PSNR. We do not have access to this oracle pattern. We want to use a risk to estimate the best pattern and to rank all patterns. Note that the ranked top-1 pattern by a risk is usually NOT the oracle pattern.

To gauge the ranking power of a risk, it is not informative to compare the rank of the oracle pattern rated by different estimators, because the ranks do no reflect reconstruction quality differences. It is also not enough to only look at the reconstruction quality difference between the top-1 pattern and the oracle pattern for two reasons. Firstly, a small quality difference can be a coincidence due to specific textures or scenes being insensitive to choices of pattern. Secondly, even a risk that rank patterns poorly may find an acceptable pattern once in a while (as illustrated in \fref{fig:score_sorted_by_risk} (d)).

Therefore, in our evaluation protocol, we propose to measure two descriptive statistics of risk estimators:
\begin{enumerate}
    \item \textbf{Average quality difference} between using the oracle and the top-$K$ patterns ranked by an estimator. This statistic evaluates the absolute reconstruction quality drop when one uses top patterns selected by a risk estimator compared to using the oracle pattern. Furthermore, if top-$K$ average difference is an increasing function of $K$, then the risk estimator likely has good ranking power on patterns. Mathematically, we define
        \begin{equation}
        \Delta_K = \frac{1}{NK} \sum_{n=0}^{N-1}\sum_{i=0}^{K-1}(\underset{\text{oracle score}}{\underbrace{s_{n}^{*}}} - \underset{\text{$i$-th score}}{\underbrace{s_{n,i}}}),
        \label{eq:top_k_diff}
        \end{equation}
        where $\Delta_K$ is the top-$K$ average difference, $s_{n}^*$ is the score of the oracle pattern on the $n$-th radiance map ($n = 0,1,\ldots,N-1$, with $N = 1494$ in this paper), $s_{n,i}$ is the score of the $i$-th top pattern as ranked by the risk.
    \item \textbf{Probability} that the reconstruction quality difference between the oracle pattern and the top-1 pattern is above certain pre-determined threshold. This statistic measures: given a threshold that one considers as critical, what is the probability that using a particular risk estimator will not yield satisfactory results. Formally, we define
        \begin{equation}
        Q(\eta) = \frac{1}{N} \sum_{n=0}^{N-1} \mathbb{I}\left\{\frac{s_{n}^{*}-s_{n,1}}{s_{n}^{*}} > \eta\right\},
        \label{eq:top_k_Q}
        \end{equation}
        where $Q$ is the probability of having a difference above a threshold $\eta$, $\mathbb{I}[\cdot]$ is an indicator function.
\end{enumerate}

\textbf{Results}. We show in Table~\ref{tab:risk-eval-results} the top-$K$ average differences and $Q$ scores at two thresholds across the pattern ranking dataset. The full SVE-Risk is capable of selecting a pattern that will yield a reconstruction with close to oracle performance, with an average $\mu$PSNR drop for the top-1 pattern around 1 dB. SVE-Risk without the neighborhood penalty term can still pick a reasonably good pattern, but is almost always subpar compared to pattern selected by full SVE-Risk. SNR-Risk as well as its variant SNR-Risk$_{\text{MSE}}$ are incapable of picking a good pattern in almost all scenarios, and equipping SNR-Risk with MSE for overflowing pixels does not help SNR-Risk. This experiment shows the significance of properly assigning surrogate risk to recoverable overflowing pixels and exposure/gain control element binding.

\begin{figure}[h]
\centering
\begin{tabular}{cccc}
\hspace{-2ex}\includegraphics[width=0.23\linewidth]{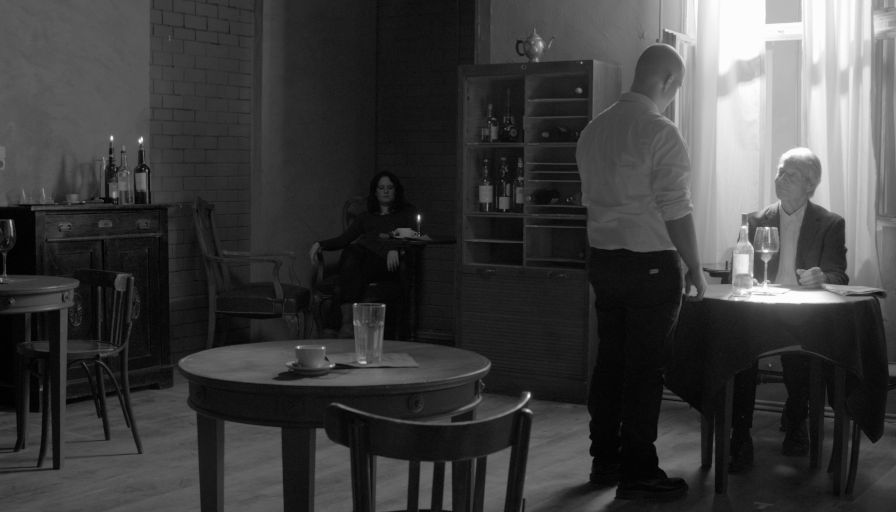}&
\hspace{-2ex}\includegraphics[width=0.23\linewidth]{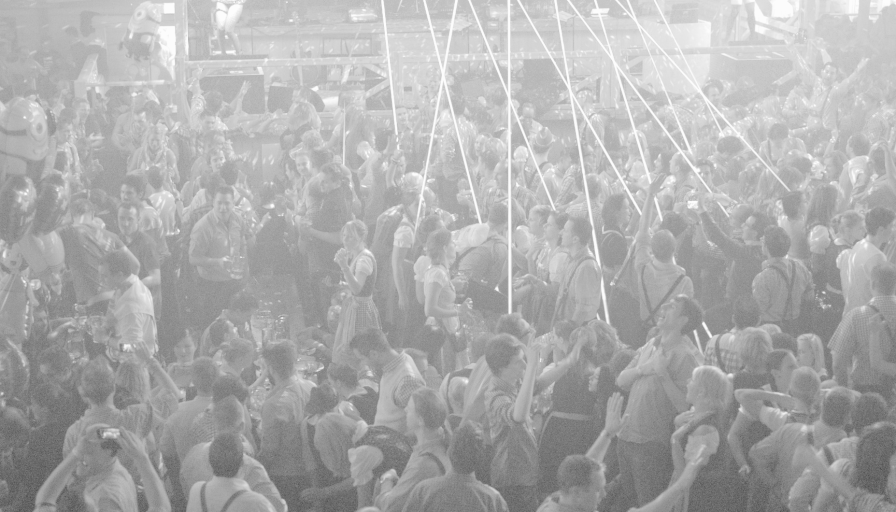}&
\hspace{-2ex}\includegraphics[width=0.23\linewidth]{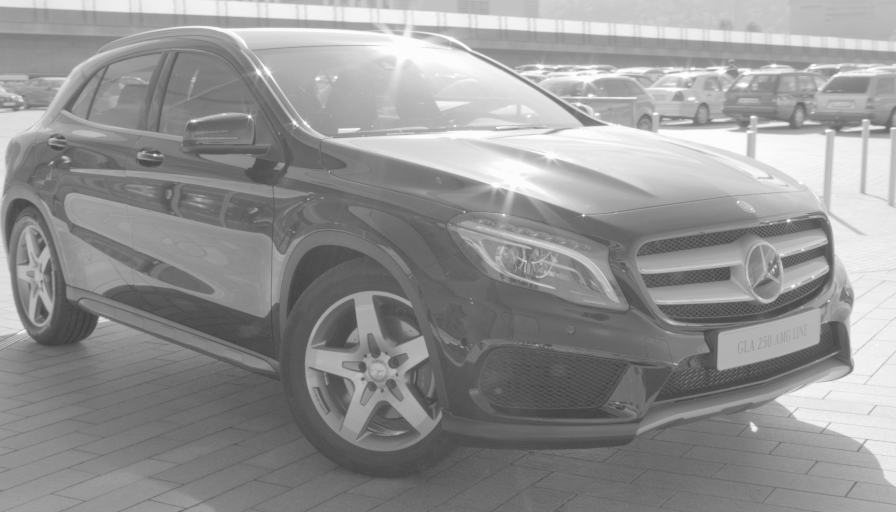}&
\hspace{-2ex}\includegraphics[width=0.23\linewidth]{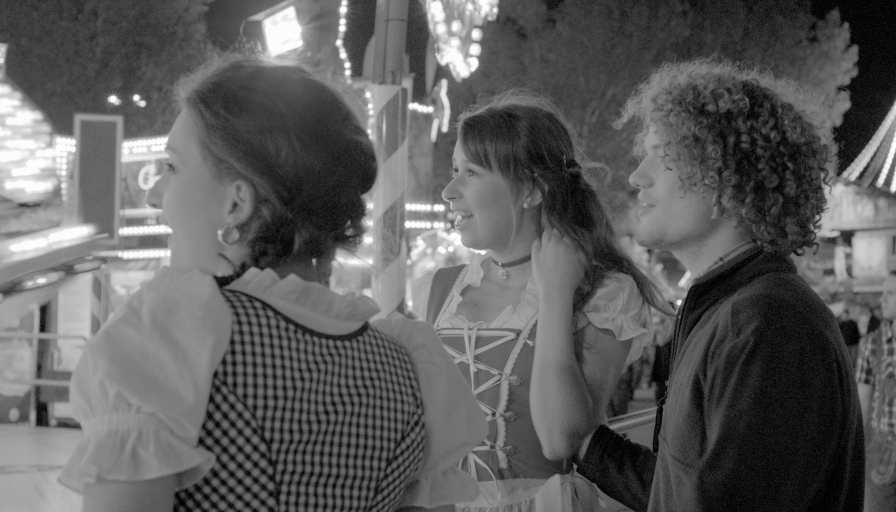}\\
\footnotesize{Scene 1} & \footnotesize{Scene 2} & \footnotesize{Scene 3} & \footnotesize{ Scene 4}
\end{tabular}

\vspace{2ex}
\scalebox{0.8}{
\begin{tabular}{clcc}
\hline\hline
\multicolumn{4}{c}{Optimal patterns (oracle scheme)}\\
\hline
Image & How bright? & Optimal exposure $\tau$ & Optimal gain $\alpha$\\
\hline
Scene 1 & All dark    & \{10,1,10,10\} & \{4,2,4,4\}\\
Scene 2 & All bright  & \{1,1,1,1\}    & \{2,1,2,2\}\\
Scene 3 & Half-half   & \{1,10,10,1\}  & \{1,2,4,2\}\\
Scene 4 & More dark   & \{1,10,10,10\} & \{1,4,4,1\}\\
\hline
\end{tabular}}
\caption{Scenes and their corresponding oracle patterns.}
\label{fig:pattern_across_scene}
\end{figure}

\begin{figure*}[!h]
\setlength\tabcolsep{1pt}
\renewcommand{\arraystretch}{0.5}
\centering
\resizebox{\textwidth}{!}{%
\begin{tabular}{ccccc}
\includegraphics[width=0.2\textwidth]{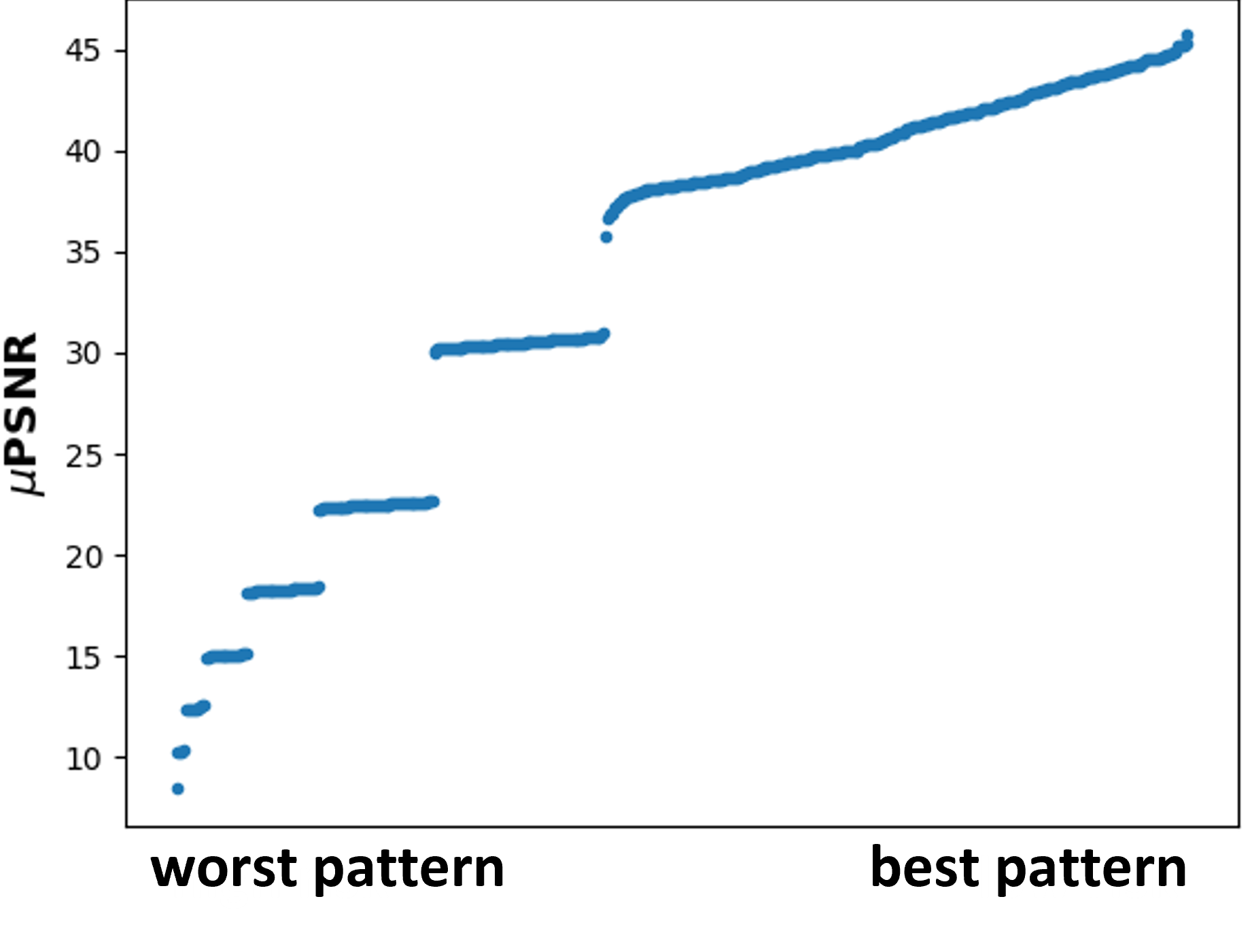}&
\includegraphics[width=0.2\textwidth]{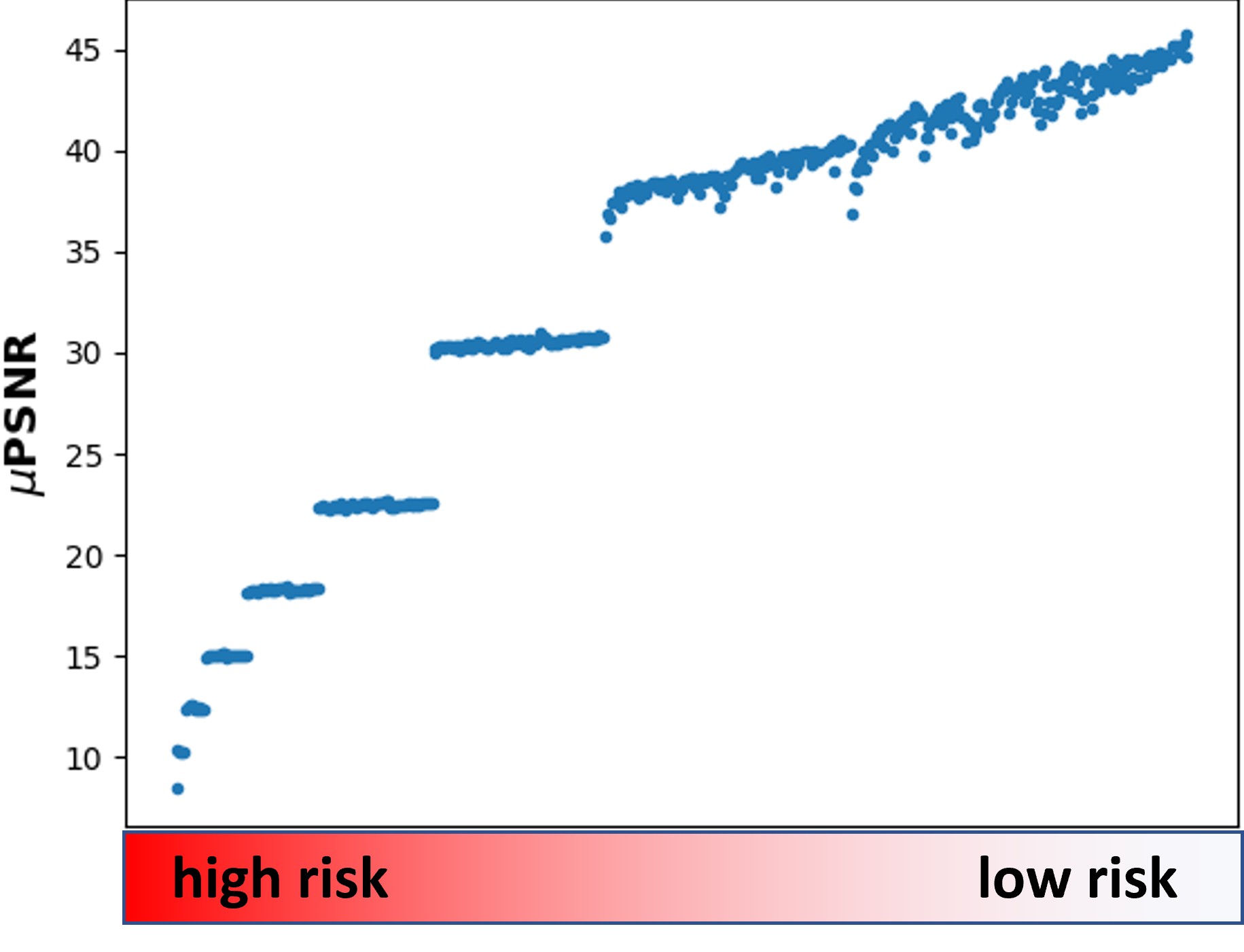}&
\includegraphics[width=0.2\textwidth]{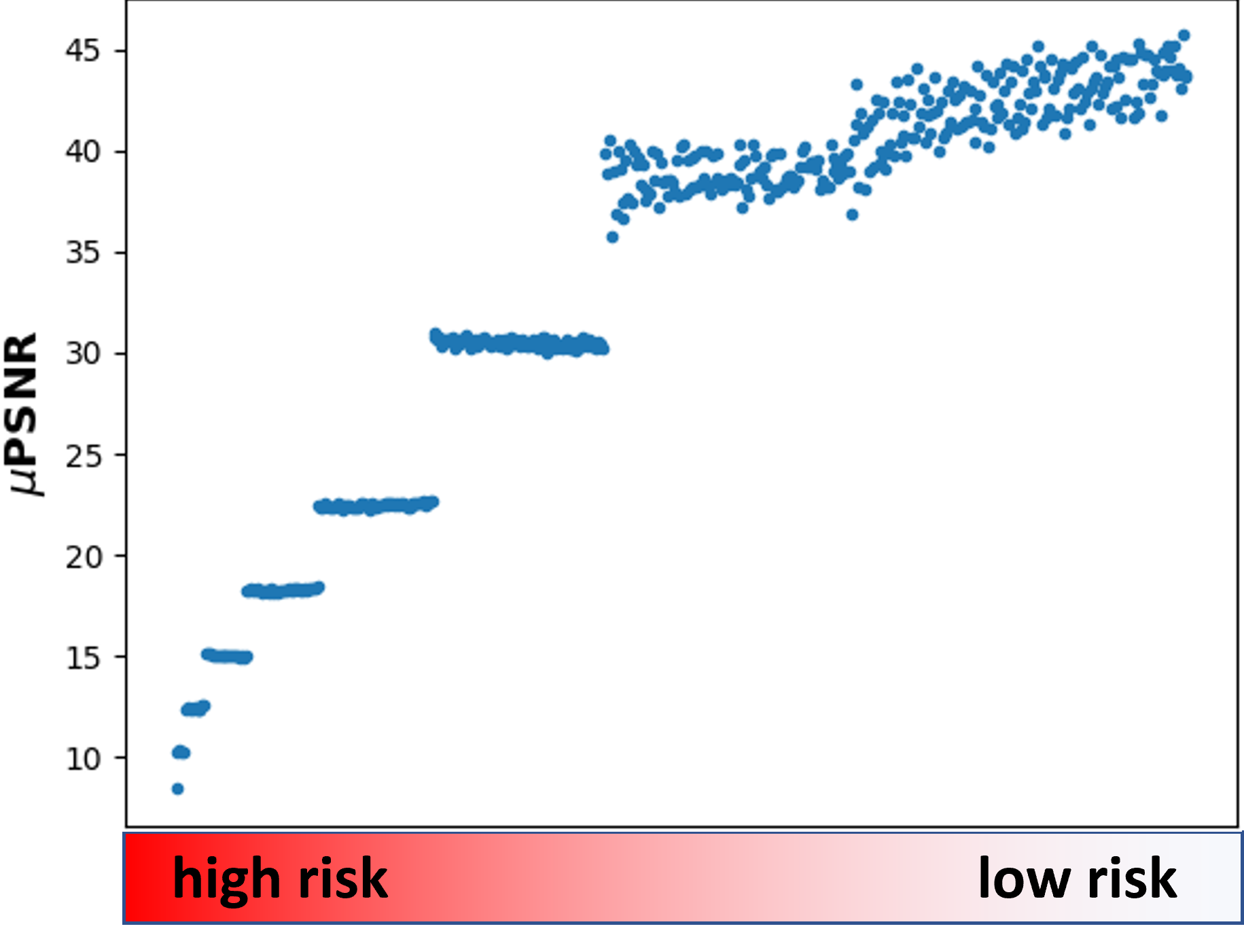}&
\includegraphics[width=0.2\textwidth]{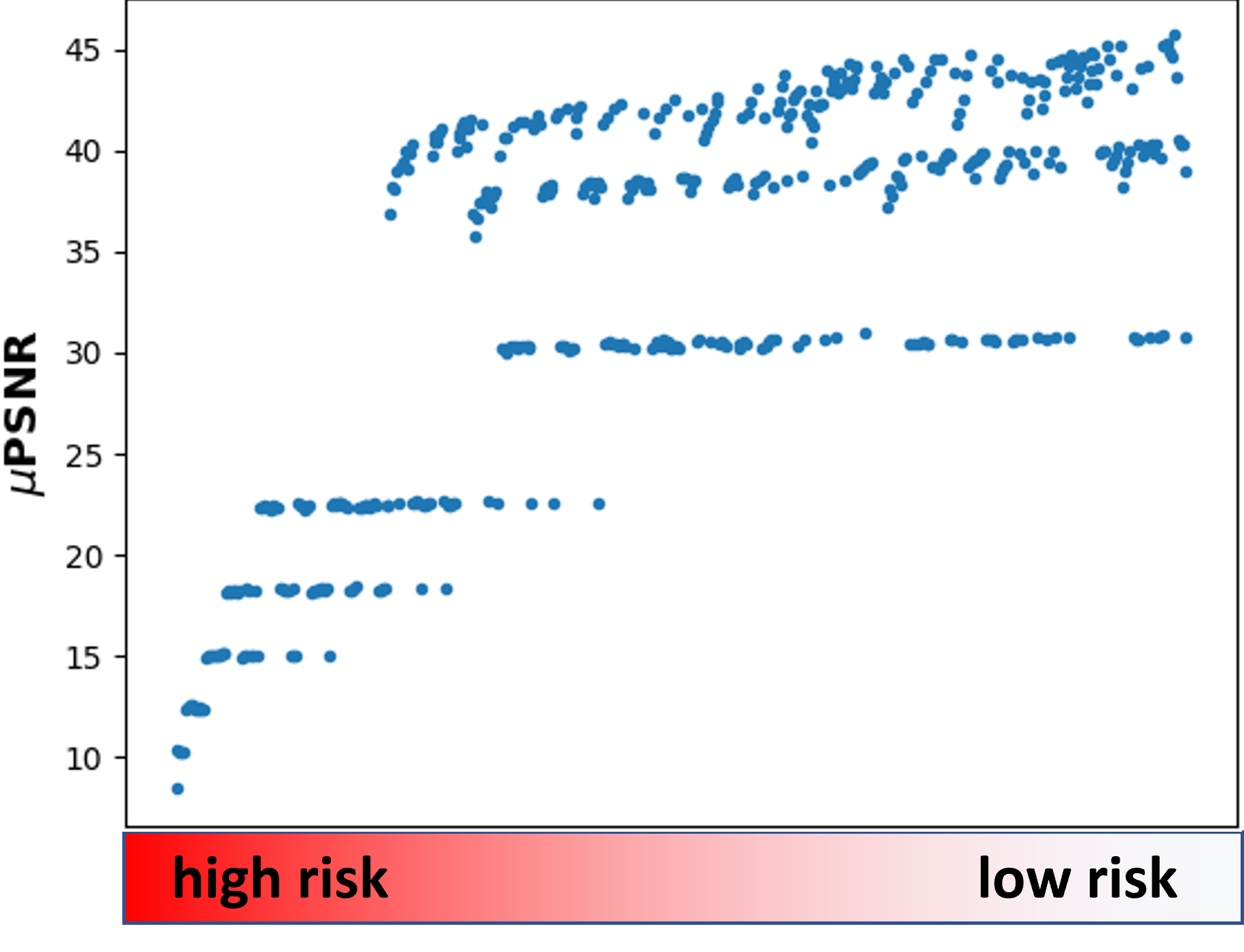}&
\includegraphics[width=0.2\textwidth]{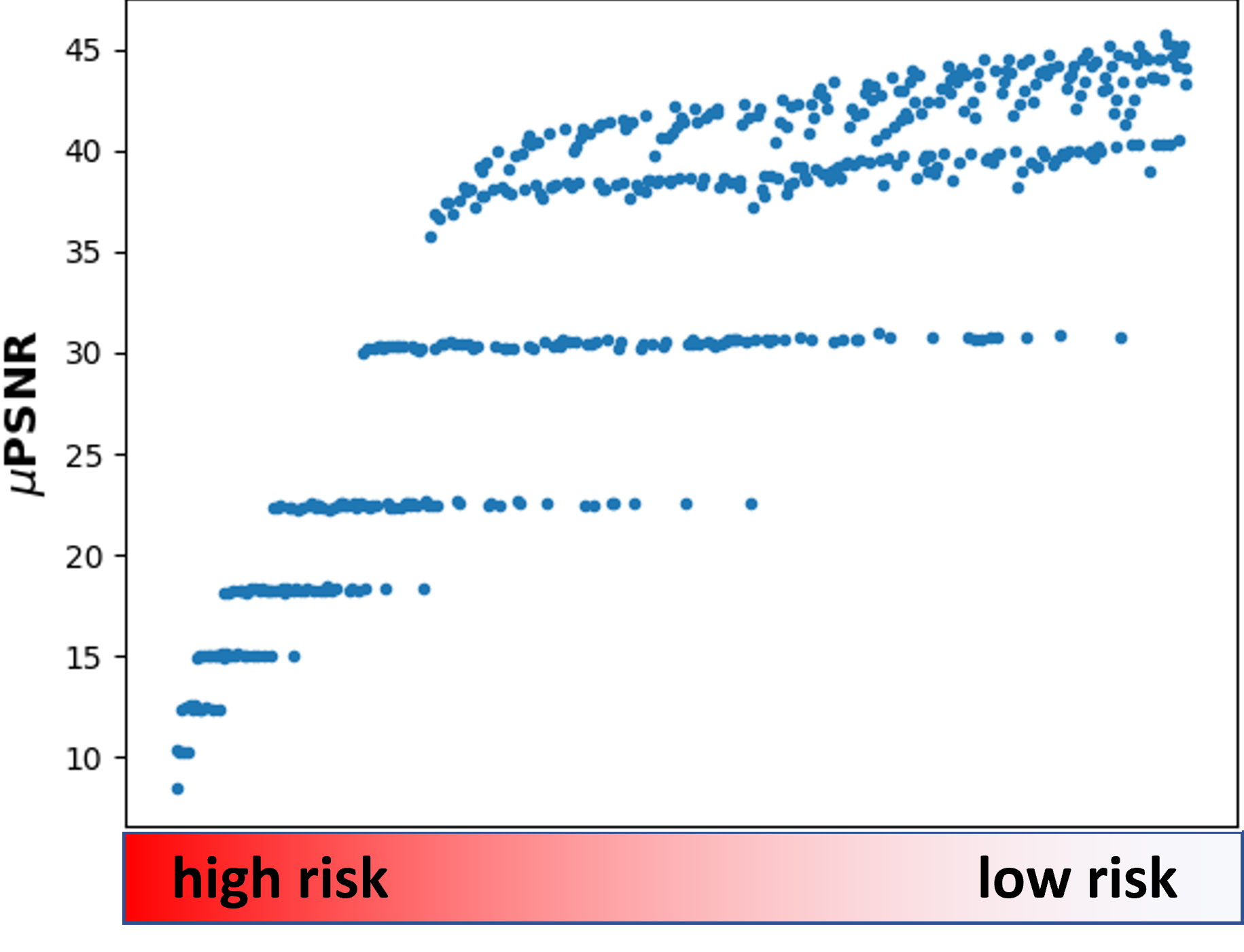}\\
(a) $\mu$PSNR & (b) SVE risk & (c) SVE$_{\text{w/o}}$ risk & (d) SNR risk & (e) SNR$_{\text{MSE}}$ risk
\end{tabular}
}
\caption{Scatter plots of $\mu$PSNR of exposure patterns with their x-coordinates sorted. Each dot represents a pattern. \textbf{(a)} Exposure patterns sorted by their corresponding $\mu$PSNR scores from low to high, representing what an \textit{ideal} risk estimator should achieve. An ideal risk estimator always assigns lower risks to patterns yielding higher $\mu$PSNR. \textbf{(b) -- (e)} Exposure patterns sorted by their estimated risk values using various risk estimators. The y-coordinate of a dot is the $\mu$PSNR of reconstruction using that pattern.  }
\label{fig:score_sorted_by_risk}
\end{figure*}

\textbf{How do the Optimal Patterns Look Like?} To give readers an idea of how the optimal exposure/gain patterns look like, we show in \fref{fig:pattern_across_scene} four randomly selected scenes and their corresponding optimal exposure/gain patterns. As the radiance of the scenes change from all dark to all bright, the optimal patterns change from all-high to all-low. This variety of scenes with the experimental results suggest that our proposed scheme is able to adaptively select the exposure and gain based on the radiance.

\begin{figure*}[!h]
\setlength\tabcolsep{1pt}
\renewcommand{\arraystretch}{0.5}
\centering
\begin{tabular}{c}
     \includegraphics[width=0.99\linewidth]{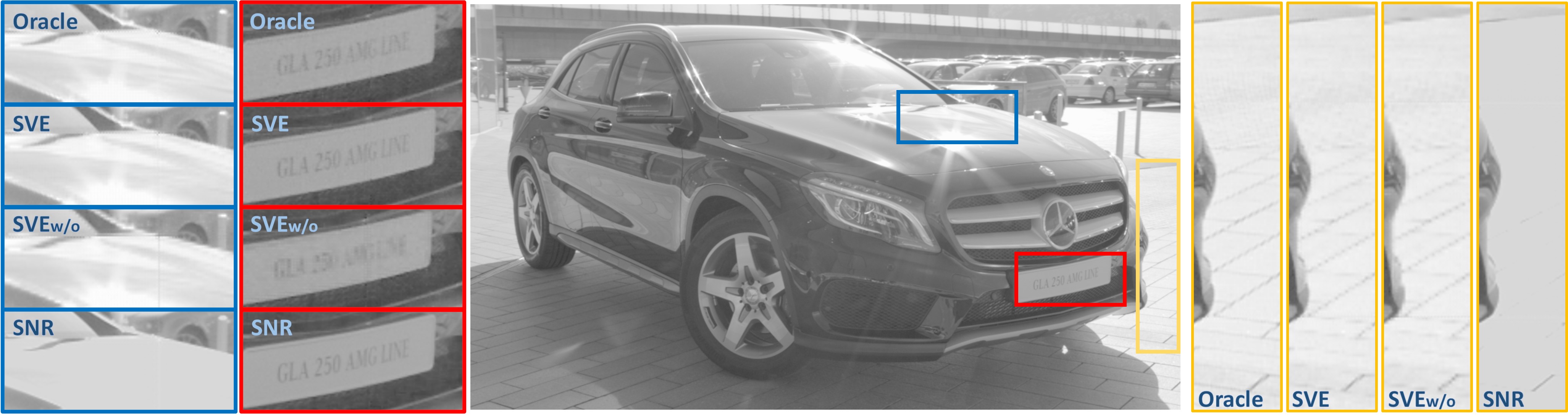}\\
     (a)\\
     \includegraphics[width=0.99\linewidth]{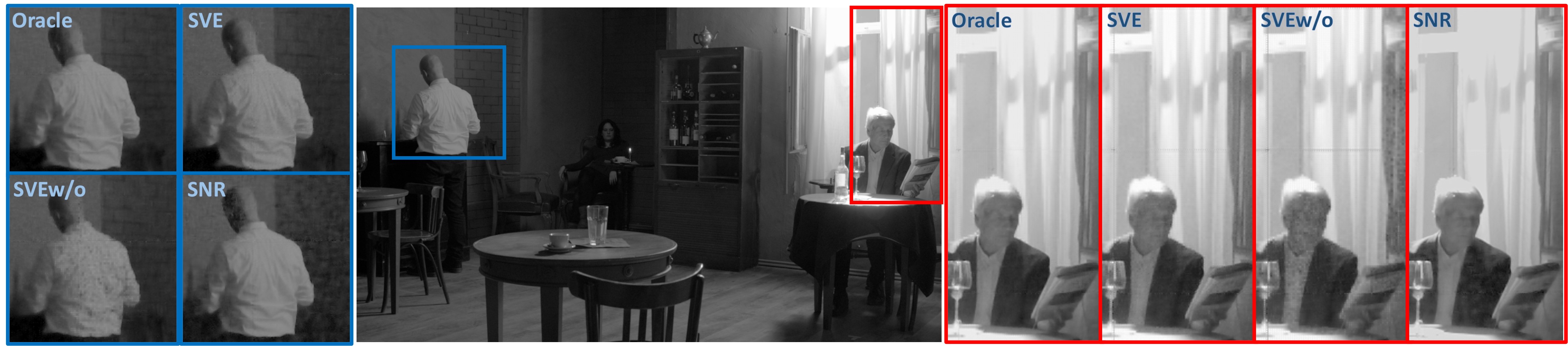}\\
     (b)\\
\end{tabular}
\caption{Qualitative comparisons of risk estimators. Images in the middle are the ground truth. Images on the sides are reconstructions using simulated readouts captured with spatially-varying exposure patterns selected by different risk estimators. Oracle denotes the best possible pattern through exhaustive search. \textbf{(a)} SNR-Risk is dominated by relatively darker regions, so it selects an exposure pattern such that the high-light textures are irrecoverable. SVE-Risk$_\text{w/o}$ fails to retain low-light details, e.g., characters on the license plate. Oracle and SVE-Risk selections preserve most information across high dynamic ranges. \textbf{(b)} SNR-Risk is unable to pick a spatially-varying exposure/gain, and therefore compromises texture in high-light regions and introduces irrecoverably heavy noise in low-light region. SVE-Risk$_\text{w/o}$ loses details in medium exposure regions such as the human face and the white shirt. Oracle and SVE-Risk selections trade off high/low flux regions differently: the oracle pattern better preserves the face but loses some details of the curtain, while SVE-Risk's top pattern preserves all curtain details but results in a slightly more blurry face. The reconstruction algorithm is Restormer for all images. All images are tone-mapped.}
\label{fig:qualitative_visual}
\end{figure*}

\textbf{Visualize Ranking Power}. To illustrate the ranking power of SNR-Risk and SVE-Risk, we show a scatter plot of $\mu$PSNR of a scene sorted by risk values in \fref{fig:score_sorted_by_risk}. We remark that this is a novel visualization of the performance, as we have not seen a similar plot in the literature.

To interpret the results of this plot, we note that the $x$-axis of the plots is the risk ranked from high to low. There are four risks: SNR, SVE, SVE$_{\text{w/o}}$ and SNR$_{\text{MSE}}$. The ideal risk for our task is $\mu$PSNR. If we use $\mu$PSNR as the metric to rank the patterns, we will have a scatter plot shown in \fref{fig:score_sorted_by_risk}(a). A better pattern ranked by $\mu$PSNR will, of course, give a higher $\mu$PSNR. When we evaluate other risks, we see that the proposed SVE-Risk has the closest behavior to the ideal risk. In contrast, SNR-based risks show an overlapped behavior. This means that if we use SNR to pick the pattern, we will not be able to tell which pattern is the best because for the same SNR ($x$-axis), we have multiple patterns on the $y$-axis.

\textbf{Staircase $\mu$PSNR Behavior}. Readers may wonder about the step-wise behavior. This is due to the image histogram, as the scene may contain large flat regions of similar radiance values. As the minimum exposure and gain go above certain thresholds, the brightest large region of scene becomes completely saturated and irrecoverable, causing significant quality drop. Such drop is intrinsic to the scene itself and related to values that local exposure and gain can take, but no pattern is guaranteed to be in any particular cluster as the scene varies (i.e., no intrinsically bad pattern).

\begin{figure*}[h]
\centering
\includegraphics[width=0.9\linewidth]{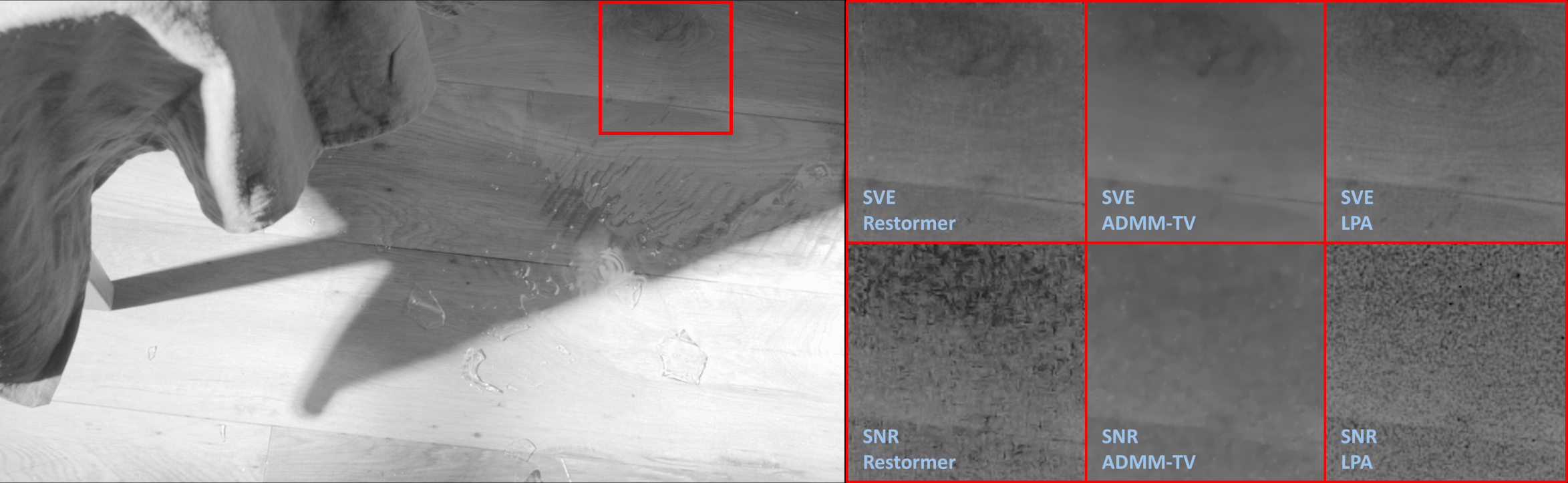}
\caption{Images reconstructed by different algorithms. Using the sub-optimal spatially-varying exposure pattern selected by SNR-Risk, all three reconstruction algorithms fail to recover the texture of the hardwood flooring.}
\label{fig:algorithms_visual}
\end{figure*}

\textbf{Visualize the Patterns}. A visualization of adopting the top pattern selected by different risks for capturing is shown in \fref{fig:qualitative_visual}. In \fref{fig:algorithms_visual}, we show an example of reconstructing simulated readouts captured with SVE-Risk top pattern and SNR-Risk top pattern using different reconstruction algorithms.

\begin{figure}[h]
\centering
\begin{tabular}{cc}
\includegraphics[width=0.45\linewidth]{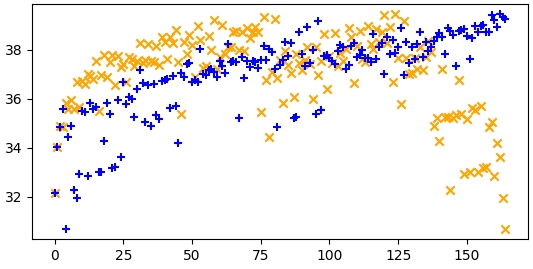}&
\includegraphics[width=0.45\linewidth]{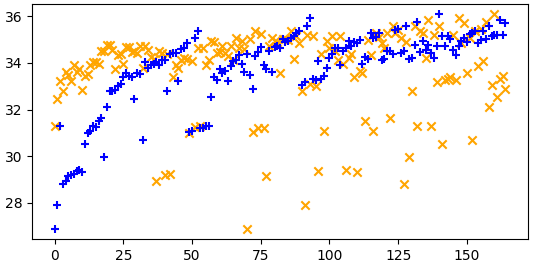}
\end{tabular}
\caption{$\mu$PSNR versus risk rank scatter plots. Even if we throw away bad pixels, SNR-Risk still cannot generate a smooth curve. In contrast, SVE-Risk produces a largely smooth curve. Color code: (\textcolor{blue}{SVE}, \textcolor{orange}{SNR})}
\label{fig: Bad pixels}
\end{figure}

\textbf{Can SNR-Risk Work if We Discard ``Bad'' Patterns?} A common question people ask is that would SNR-Risk perform better if we throw away the bad patterns. Our answer is no. Firstly, we simply cannot discard bad patterns when there is no intrinsically bad pattern. Secondly, even if we analyze current scene and discard all patterns that may yield large saturated region, pixel-wise SNR will still not use nonuniform exposure levels. This can be seen in the $\mu$PSNR v.s. risk rank scatter plots as shown \fref{fig: Bad pixels}.

\subsection{Verify the Universality of Patterns}
\label{sec:experiments:empirical}

In this subsection, we describe our discovery that an exposure/gain pattern is universal for \emph{many} image reconstruction algorithms. This is a significant departure from the recent trend of camera-algorithm co-optimization where people have been arguing that jointly optimizing the pattern and the algorithm is essential. Our experiments in this subsection show the opposite. We find that the design of the exposure/gain pattern can be completely decoupled from the design of the image reconstruction algorithm.

\textbf{Spearman's ranking correlation}. To evaluate the dependency of the pattern and the algorithm, we need some notion of correlation between the two factors. The metric we consider in this paper is the Spearman's ranking correlation \cite{spearman}, although other types of correlations can also be used.

For each ground truth radiance image $\vtheta$, we synthesize 495 raw sensor readouts $\{\vx_0, ..., \vx_{494}\}$, one for each pattern equivalence class. For each readout, we use three distinct algorithms $\{f_0, f_1,f_2\}$ to reconstruct the radiance image and evaluate the reconstruction quality using three different metrics $\mathcal{M}_{\mu\text{PSNR}}, \mathcal{M}_{\mu\text{SSIM}}, \mathcal{M}_{\mu\text{LPIPS}}$. We define a score as
\begin{equation}
s_{i,j,k}=\mathcal{M}_k(f_j(\vx_i),\vtheta),
\end{equation}
where $s_{i,j,k}$ denotes the score using $i$th pattern, $j$th algorithm, and metric $k$. The exhaustive evaluation results are collated to create a pattern ranking data set. We assess whether a monotonic relationship exists between a pair of reconstruction algorithms as pattern varies by computing Spearman's ranking correlation coefficient \cite{spearman} over scores of a metric
\begin{align*}
    &\rho_{j,j^\prime,k} =
    \text{Spearman}(\{(s_{i,j,k},s_{i,j^\prime,k})
    \mid i = 0, \cdots , 494 \}) ,
\end{align*}
where a tuple of scores $(s_{i,j,k},s_{i,j^\prime,k})$ is treated as an observation, associated with which a p-value describes the probability that no monotonic relationship exists between them. The correlation coefficients are computed for every pair of algorithms over all 1494 images from NTIRE dataset.

\begin{table*}[!t]
\caption{Average (standard deviation) and median correlation coefficients and average p-value of each correlation coefficient across NTIRE dataset}
\resizebox{\textwidth}{!}{%
\begin{tabular}{c|ccc|ccc|ccc}
\hline
\hline
 & \multicolumn{3}{c|}{$\mathbf{\mu PSNR}$} & \multicolumn{3}{c|}{$\mathbf{\mu SSIM}$} & \multicolumn{3}{c}{$\mathbf{\mu LPIPS}$}\\
 & average $\rho$ & median $\rho$ & average p-value & average $\rho$ & median $\rho$ & average p-value & average $\rho$ & median $\rho$ & average p-value \\
\hline
ADMM-TV v.s. LPA & $0.937 \pm 0.035$ & $0.952$ & $<10^{-7}$ & $0.805 \pm 0.116$ & $0.830$ & $<10^{-7}$ & $0.740 \pm 0.195$ & $0.775$ & $0.008 \pm 0.079$ \\
LPA v.s. Restormer & $0.877 \pm 0.070$ & $0.878$ & $<10^{-7}$ & $0.826 \pm 0.053$ & $0.826$ & $<10^{-7}$ & $0.773 \pm 0.112$ & $0.770$ & $<10^{-7}$ \\
Restormer v.s. ADMM-TV & $0.891 \pm 0.042$ & $0.885$ & $<10^{-7}$ & $0.701 \pm 0.130$ & $0.719$ & $<10^{-7}$ & $0.564 \pm 0.306$ & $0.625$ & $0.071 \pm 0.252$ \\
\hline\hline
\end{tabular}%
}
\label{tab:empirical}
\vspace{-2ex}
\end{table*}

\begin{figure*}[!t]
\setlength\tabcolsep{1pt}
\renewcommand{\arraystretch}{0.5}
\centering
\begin{tabular}{c}
\includegraphics[width=1\linewidth]{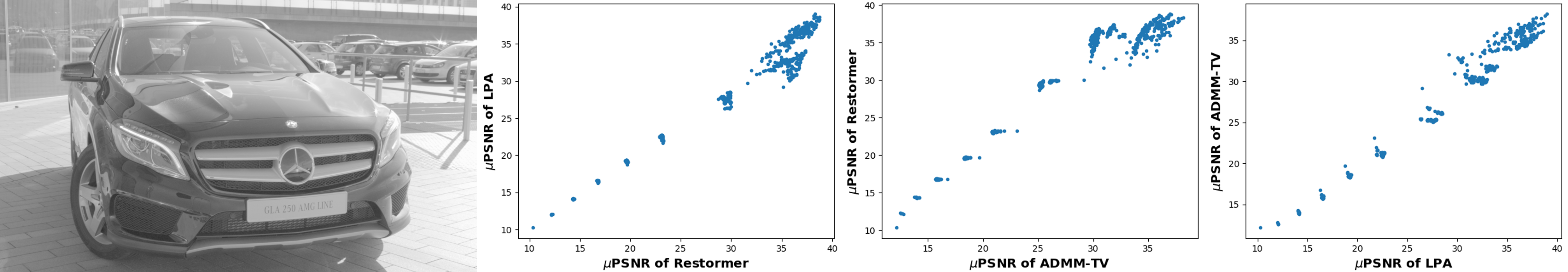}\\
\end{tabular}
\caption{Tone-mapped ground truth HDR scene and cross comparisons between reconstruction algorithms in $\mu$PSNR. Each dot in the scatter plots represents an exposure pattern. The x and y coordinates are $\mu$PSNR scores of the reconstructions by different algorithms on the image captured using that exposure pattern. The average correlation coefficient of the three scatter plots is approximately $\rho = 0.86$.}
\label{fig:corr_scatter}
\end{figure*}

\textbf{Experiment Protocol}. The overall procedure of the experiments is as follows. Given a scene radiance map, we exhaustively evaluate reconstruction quality of distinct reconstruction algorithms on synthesized sensor readouts for all patterns returned by our enumeration algorithm. Taking reconstruction quality scores as data samples, we calculate Spearman's ranking correlation coefficient for every pair of reconstruction algorithms. We conduct hypothesis testing
\begin{itemize}

\item $\mathcal{H}_0:$ the reconstruction quality between two algorithms are uncorrelated as exposure pattern varies,
\item $\mathcal{H}_a:$ the reconstruction quality are positively correlated.
\end{itemize}
This procedure is repeated on every sample of NTIRE \cite{NTIRE} dataset. We report the average and median of correlation coefficients across 1494 images for each metric in Table~\ref{tab:empirical}.

\textbf{Results}. To give a sense of the reported value, we also show scatter plots of $\mu$PSNR of pairs of reconstruction algorithms in \fref{fig:corr_scatter}, with detailed numbers shown in Table~\ref{tab:empirical}.

The three scattered plots in \fref{fig:corr_scatter} are worth discussing. These three subplots are the $\mu$PSNR comparison between LPA, ADMM-TV, and Restormer. The scattered plot shows a surprisingly strong correlation between any pair of the methods: If a $2\times 2$ pattern favors LPA, it also favors ADMM-TV, and similarly for other pairs. Therefore, at least based on this limited set of experiments, we find that if a pattern is good, it is good for all reconstruction algorithms; if it is bad, it is bad for all reconstruction algorithms. By inspecting the numbers in Table~\ref{tab:empirical}, we further note that the Spearman's correlation coefficients are all in the range of 0.87 or above (for $\mu$PSNR). For other evaluation metrics $\mu$SSIM and $\mu$LPIPS, we also see a high correlation coefficient.

We believe that this finding is new and perhaps less expected. The implication is that if we need to design the multiplexing pattern, there is no need to consider the image reconstruction algorithm. This is a good news from the point of a designer's perspective. Co-optimization is not always preferred because we do not want the patterns to be dependent on a particular algorithm. If we can modularize the designs of the two, the debugging and analysis of the methods will be significantly easier.

\subsection{Real Experiments}
\label{sec:experiments:real}
In this section, we test the SVE-Risk and SNR-Risk on real camera raw readouts and show the feasibility of the proposed risk on real hardware for exposure pattern selection. Since no SVE sensor is available to us, we interlace real camera raw readouts to synthesize images captured with SVE patterns.

\subsubsection{Experiment Settings and Procedure}
\newcommand{\thecamera}[0]{Sony Alpha 7 II camera}
For each high dynamic range scene, we use a \thecamera{} to capture 9 differently exposed LDR frames. We capture 5 HDR scenes in total. For ease of camera parameter calibration, we keep the camera ISO at 100 and gradually increased the exposure time from 1/80 sec to 3.2 sec, doubling from one frame to the next. Our imaging model based risks require knowledge of dark current and read noise level. For these two parameters, we estimate them by capturing three dark frames (ISO 100, exposure 1/80 sec, 0.2 sec, 1.6 sec). We generate a pseudo ground truth radiance map by fusing all 9 frames. We synthesize all possible SVE captures using following procedure:
\begin{enumerate}
    \item Enumerate all possible SVE patterns with 9 different exposures.
    \item For each pattern, pick the corresponding frames from the 9 raw frames.
    \item Interlace picked frames to generate an SVE frame by taking every other pixel.
\end{enumerate}
Then, we use our trained Restormer to reconstruct scene radiance for every SVE captures and compare the reconstruction against the pseudo ground truth. We build the empirical radiance histogram of a scene by treating the SVE capture with exposure (1/80 sec, 1/20 sec, 1/5 sec, 4/5 sec) as the pilot frame. We use this histogram to calculate SVE-Risk for every pattern. For SNR-Risk, we used the pseudo ground truth of a scene for calculation.

The camera response function (CRF) of \thecamera{} raw readout is close to linear at ISO 100 except when the photon charge accumulated is near full well capacity. Therefore, we use a linear CRF and aggregate the entire conversion from charge to final ADC readout (voltage follower gain, column amplifier gain, and output amplifier gain) into conversion gain term $\alpha$ in \eref{eq: main sensor model}. The final parameters used in this experiment are listed in table \ref{tab:real_exp_parameters}.

\begin{figure*}[t]
\setlength\tabcolsep{1pt}
\renewcommand{\arraystretch}{0.5}
\centering
\begin{tabular}{c}
     \includegraphics[width=0.99\linewidth]{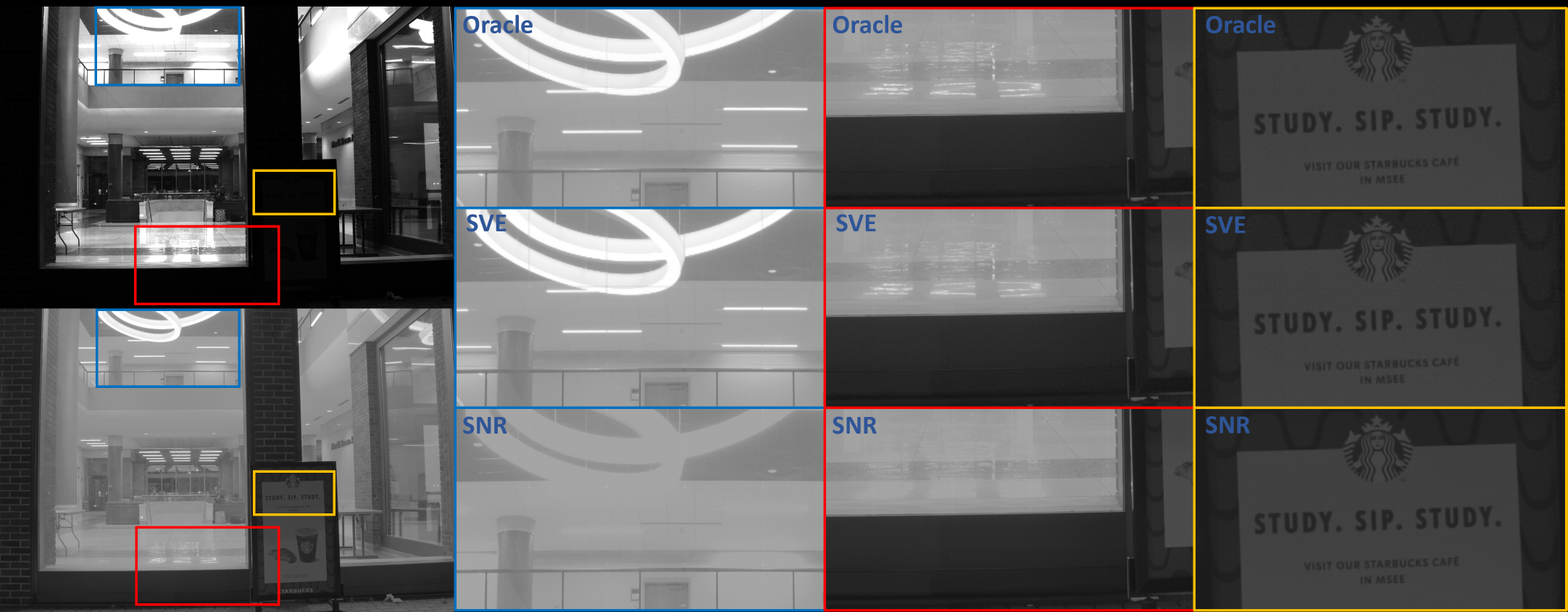}\\
     (a) \\
     \includegraphics[width=0.778\linewidth]{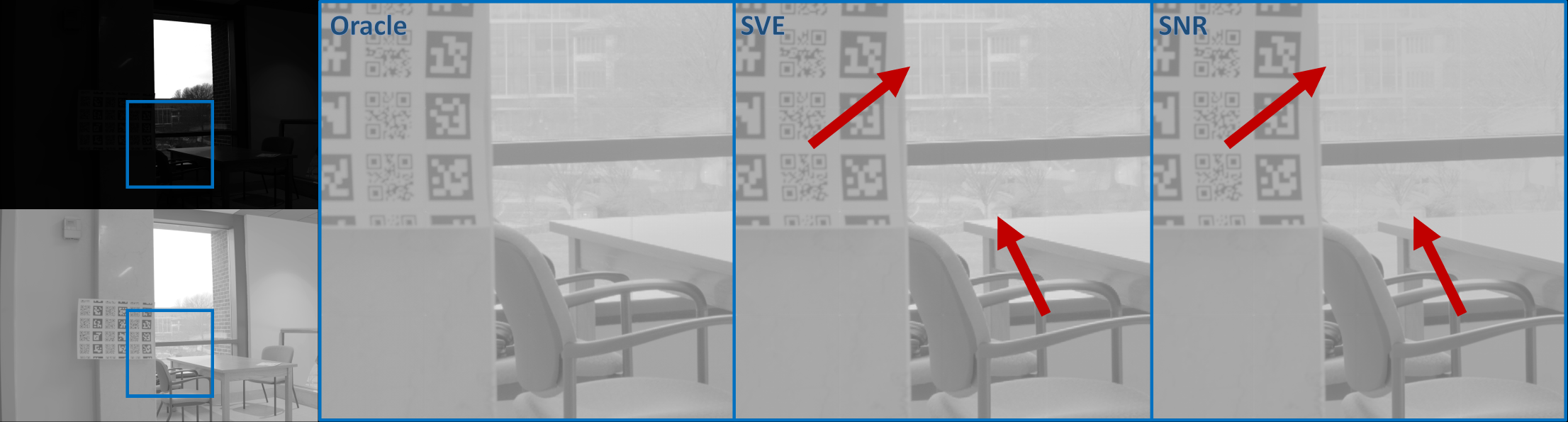}
     \includegraphics[width=0.21\linewidth]{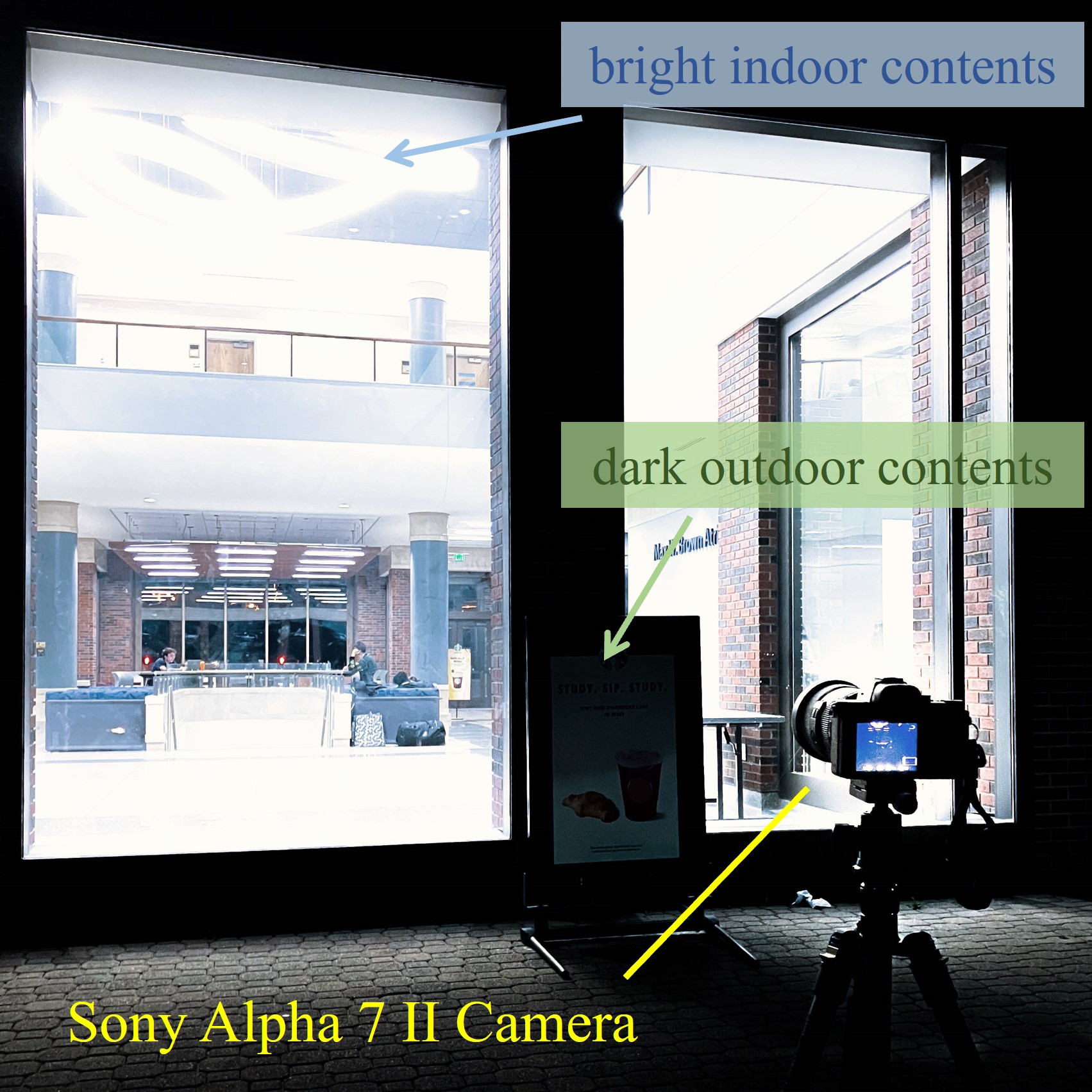}\\
     \hspace{0.3\linewidth} (b) \hspace{0.45\linewidth} (c)\\
\end{tabular}
\caption{Reconstruction results on real images and experimental setup. Input images with spatially-varying patterns are generated by interlacing raw images captured at various exposures. For \textbf{(a)} and \textbf{(b)}, we show full frames of pseudo ground truth in both linear scale and tone-mapped scale (top-left and bottom-left, respectively). The linear scales are clipped at 95 percentile radiance level and are normalized to $[0, 255]$. We also show zoomed-in regions of tone-mapped reconstructions on the best possible spatially-varying pattern via exhaustive search (oracle), the best pattern selected by SVE-Risk (SVE), and the best by SNR-Risk (SNR). The reconstruction algorithm is Restormer for all images. \textbf{(a)} The texts and textures at low light shown in the orange box are preserved equally well in all three reconstructions, but SNR-Risk reconstruction loses all details of high-light regions. \textbf{(b)} The dark regions (QR code, wall, etc.) of all three reconstructions are indistinguishable, but SNR-Risk reconstruction loses outdoor high-light details (red arrows). \textbf{(c)} Experimental setup. We capture real images of HDR scenes using a Sony Alpha 7 II camera. }
\label{fig:real_visual}
\end{figure*}

\subsubsection{Results}
We show the quantitative reconstruction quality in terms of $\mu$PSNR in table \ref{tab:real_quant} and two sets of qualitative results in \fref{fig:real_visual}. As evidenced in both quantitative and qualitative results, using top pattern selected by SVE-Risk can yield near-optimal final reconstruction.

\begin{table}[h]
\caption{Measured parameters in terms of equivalent charges for real camera experiment}
    \centering
\scalebox{0.85}{
    \begin{tabular}{ll}
    \hline
    \hline
         Gain $\alpha$                      & 5.25 e- per ADC Unit \\
         Exposure $\tau$                    & 1/80 sec, 1/40 sec, ..., 1.6 sec \\
         Read noise $\sigma_{\text{read}}$  & 1.3 e- \\
         Camera response function           & Linear \\
         dark current                       &1.4e-/sec\\
         Clip threshold                     &84667e-\\
    \hline
    \end{tabular}
    }
    \label{tab:real_exp_parameters}
\end{table}

\begin{table}[h]
    \centering
    \caption{$\mu$PSNR of images reconstructed from raw images interlaced according to best possible patterns (oracle), patterns selected by SVE-Risk, and patterns selected by SNR-Risk, compared to the pseudo ground truth}
\scalebox{0.85}{
    \begin{tabular}{c|ccccc}
    \hline
    \hline
         & scene 1 & scene 2 & scene 3 & scene 4 & scene 5\\
    \hline
    oracle & 44.5 & 44.4 & 35.5 & 40.3 & 49.5\\
    SVE-Risk top-1 & 44.5 & 42.3 & 33.9 & 38.8 & 49.3\\
    SNR-Risk top-1 & 41.8 & 38.1 & 30.0 & 27.5 & 47.6\\
    \hline
    \end{tabular}
}
    \label{tab:real_quant}
\end{table}

\section{Conclusion}
In this paper, we report two findings about the design of a spatially varying exposure multiplexing scheme. Firstly, we show that the pixel-wise SNR is a poor metric to quantify the performance of a multiplex pattern because it fails to differentiate the recoverable cases and the non-recoverable cases. We circumvent the difficulty by proposing the SVE-Risk. Our experiments show that the pattern ranking provided by the SVE-Risk correlates extremely well with the ideal ranking. Secondly, through a large-scale experiment, we find that for spatially-varying-exposure imaging with tiled exposure patterns, it is not necessary to design a pattern selection algorithm tailored for specific reconstruction algorithm; the margin of improvement for using tailored/co-designed pattern selection algorithm is limited. Our finding is a significant departure from recent work in computational photography that advocates for sensor-algorithm co-optimization. To sensor designers, this could be good news because sensor-algorithm co-design is significantly more costly for production. However, our bigger hope is that this counterexample can stimulate more discussions about the necessity of sensor-algorithm co-optimization, and under what context would it become beneficial \emph{not} to co-optimize.

\bibliography{references}
\bibliographystyle{ieeetr}

\end{document}


\maketitle
\thispagestyle{empty}

This supplementary document contains following:
\begin{itemize}
    \item More details on parameters used for synthesis and reconstruction algorithms (\ref{sec:details})
    \item More details on counting of number of non-saturated pixels within a neighborhood (\ref{sec:counting}) 
    \item Derivation for pattern equivalence class size and pseudo-code for efficient enumeration algorithm (\ref{sec:enum})
\end{itemize}

\section{More Experiment Details}
\label{sec:details}
We describe in more detail the parameters used for the data synthesis and reconstruction algorithms in this section.

\subsection{Synthesis Parameters}
Exact parameters used for synthesis is shown in table \ref{tab:synthesis_params}

\begin{table}[h]
\centering
\begin{tabular}{lll}
\hline
Symbol & Meaning & Values Used for Experiment\\
\hline
ADC & Analog-to-Digital Converter  & 10-bit \\
& ADC least significant bit resolution & 8 e- at lowest gain, 0.1 e- at highest gain\\
& ADC lower bound & 1 e- at lowest gain, 0.0125 e- at highest gain\\
& ADC upper bound & 8185 e- at lowest gain, 102 e- at highest gain\\
Clip& full well limit & 8200 e- \\
$\tau$ & exposure of individual pixel     & \{$\times0.25,\, \times0.5,\, \times1$\} global exposure duration\\
& global exposure & 30ms\\
$\mu_{\text{dark}}$ & dark current & 0.2 e- at shortest exposure\\
QE  & quantum efficiency & 80\%\\
$\alpha$ & conversion gain & \{1, 10, 80\} (roughly correspond to ISO 100, 1000, 8000)\\
$\sigma_{\text{read}}$ & read noise & 20 e- at lowest gain, 0.25 e- at highest gain\\
CRF & Camera Response Function & Linear\\
\hline
\end{tabular}

\caption{Synthesis parameters}
\label{tab:synthesis_params}
\end{table}
Note that 
\begin{itemize}
    \item wherever applicable, values shown are mapped from their original domain to number of electrons held in the potential well of a pixel for ease of understanding.
    \item the spatially-varying-exposure can be achieved by either adjusting pixel-wise exposure duration or modulating light right on top of the sensor array (sensor side modulation). In our experiment, we assumed using pixel-wise exposure duration control.
    \item CRF is in general non-linear on most modern cameras, but since our data is synthesized and a non-linear CRF only uniformly extends the dynamic range of all pixels, we adopt a linear CRF for simplicity
\end{itemize}

\subsection{Reconstruction Algorithm Parameters}
Parameters of LPA\cite{LPA}/ADMM-TV\cite{pnp_admm} are tuned by performing grid search on HDR-Eye data set. Model of Restormer\cite{Zamir2021Restormer} is trained on HDR-Eye\cite{hdr-eye} data set and hyper-parameters are set by evaluating trained model on SIGGRAPH17\cite{siggraph17} data set.
\begin{itemize}
    \item \textbf{LPA} For LPA algorithm, we estimate each pixel with a window of size 7-by-7 and a Gaussian radial basis function with scale parameter of 1.
    \item \textbf{ADMM-TV} For ADMM-TV, we use regularization coefficient of 1 and run at most 30 iterations for reconstruction. We use scikit-image Chambolle total variation denoising implementation with default parameters for prior step.
    \item \textbf{Restormer} We use original author's implementation\footnote{https://github.com/swz30/Restormer} with following hyper-parameters $"inp\_channels": 1, "out\_channels": 1, "dim": 48, "num\_blocks": [4, 6, 6, 8], "num\_refinement\_blocks": 4, "heads": [1, 2, 4, 8], "ffn_expansion_factor": 2.66, "bias": false, "LayerNorm_type": "WithBias", "dual\_pixel\_task": false$. Since the network consumes a gigantic amount of memory even at inference time, we train the network on a fixed spatial resolution of 128-by-128. During test time, we partition image to 128-by-128 patches, reconstruct each single patch, and finally stitch patches back to original image.
\end{itemize}

\section{Counting Non-saturated Neighbors for Computing SVE-Risk}
\label{sec:counting}
\begin{figure}[!h]
    \centering
    \includegraphics[width=1\linewidth]{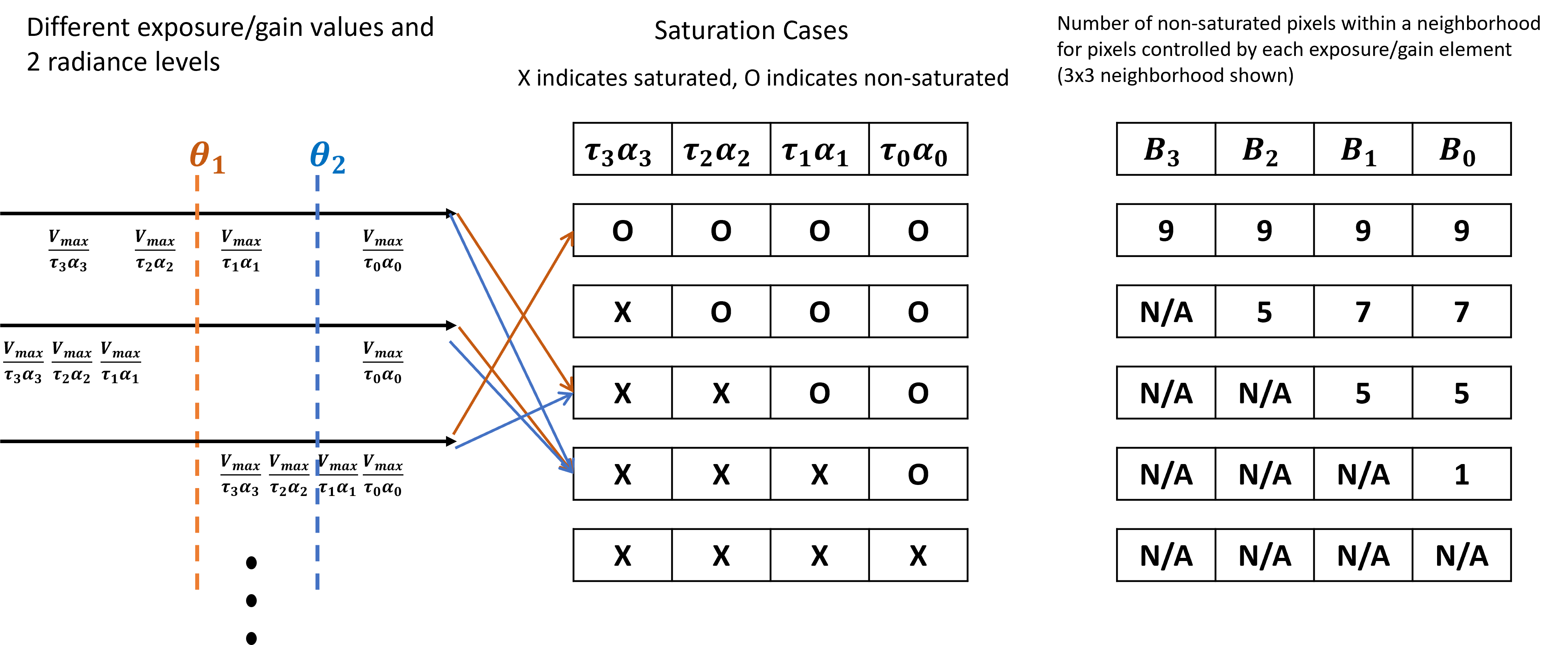}
    \caption{Example of counting number of non-saturated pixels with 3x3 neighborhood. The radiance levels and exposure/gain setting determines the saturation case, but for each case, the number of saturated pixels is fixed. Note that this is only true under our small neighborhood and pixel variation assumptions. We do not care about number of non-saturated pixels in a neighborhood for saturated exposure/gain element because saturated pixels' risks are replaced with their neighbors' and the number is not used.}
    \label{fig:counting}
\end{figure}

Under the assumptions of using small neighborhood (e.g. 3x3, 5x5) for computing SVE-Risk as well as adjacent pixels having similar values, given our knowledge about how exposure/gain pattern is tiled across the entire sensor array, we can count number of non-saturated pixels within the neighborhood of any pixel at any radiance level before we calculate risk. In fact, we don't even need the knowledge of radiance distribution or the exact exposure/gain value for the counting. 

As illustrated in figure \ref{fig:counting}, given maximum reference voltage of ADC ($V_{max}$) and exact exposure/gain values (${\{\vtau_l\},\{\valpha_l\}}$), a radiance level is mapped to a saturation case. This mapping is dependent on the exact values of $\theta$ as well as ${\{\vtau_l\},\{\valpha_l\}}$ and cannot be known beforehand. However, the mapping from saturation case to number of non-saturated pixels within a pre-specified neighborhood is fixed. Therefore, once we determine the neighborhood size that we are gonna use for computing the SVE-Risk, we can do the counting for every saturation case and each exposure/gain control element, and then construct a hash table for usage when we calculate risk for a specific scene.

Numbers counted in this way are good approximations to the ground truth numbers for majority part of an image, and only become inaccurate around thresholds $\frac{V_{max}}{\tau_l\alpha_l}$ and dark-bright region boundaries.

\section{Enumeration Size Derivation and Implementation}
\label{sec:enum}
Given an SVE array with $L$ unique exposure/gain levels (9 in our experiment) and $M$ independently configurable exposure/gain control elements (4 in our experiment), we want to evaluate risk on canonical patterns of each pattern equivalence class. To do this, we need a method to enumerate these equivalence classes. The most straight forward, albeit naive, approach is to enumerate all patterns and for each pattern, check whether another pattern within the same class has appeared before it. This implementation requires at least $L^M$ enumerations and comparisons. Although one could carry out the naive approach offline and cache equivalence classes for later usage, we show here that there is a much more elegant and efficient approach. The efficient approach comes naturally as we derive equation for calculating number of equivalence classes.

Two patterns are equivalent if and only if they are permutations of each other. This means that $M$ exposure/gain control elements are indistinguishable, while $L$ exposure/gain levels are distinguishable. The question of "how many equivalence classes are there" can therefore be rephrased as "how many ways are there to assign $M$ indistinguishable elements to $L$ distinguishable levels". 

We can assign $M$ elements to 1 level (i.e. all exposure/gain control element use same value), 2 levels, or up to $min(M, L)$ different levels. Apparently, assigning $M$ elements to 1 level is guaranteed to be non-redundant compared to assigning $M$ elements to 2 levels, if we ensure that each level chosen receive at least 1 element. It is also easy to see that given a subset of chosen levels, if we replace any level within subset with another level outside of the subset, the possible assignments generated by old subset is guaranteed to be non-redundant as compared to assignments generated by new subset. 

Now, suppose we have decided to use $k$ levels and each level must be assigned at least one element, then there are $\binom{L}{k}$ unique combinations of levels, and each combination guarantees generated assignments are non-redundant. How many non-redundant assignments are there for each combination? Well, imagine that $M$ elements are in a list, then the question is same as how many ways there are to partition the list into $k$ non-empty sub-lists, which is $\binom{M-1}{k-1}$. Therefore, we have $\binom{L}{k} \binom{M-1}{k-1}$ unique assignments if we use $k$ levels. Summing $k$ from 1 to $min(M, L)$ yields total number of unique assignments.

Following the counting process, we obtain an algorithm that directly enumerate on equivalence classes. We show pseudo-code for an implementation in algorithm \ref{alg:enum}, \ref{alg:nchoosek}

\begin{algorithm}
\caption{Pattern equivalence class enumeration}\label{alg:enum}
\begin{algorithmic}[1]
\Procedure{enumerate}{$Levels, M$}\Comment{Enumerate pattern equivalence classes given allowed exposure/gain $Levels$ and $M$ exposure/gain control elements}
\State $locations \gets [0,...,M-1]$\Comment{split location for partitioning exposure/gain control elements list}
\For{$k\gets 1,...,min(L,M)$} \Comment{Use $k$ of $L$ levels}
\State $indices \gets [0,...,k-1]$\Comment{index into exposure/gain level values}
\State $loc\_indices \gets [0,...,k-1)$\Comment{index into $locations$}
\For{$selected \gets NCHOOSEK(k, indices, Levels)$}\Comment{retrieve k levels to use}
\For{$splits \gets NCHOOSEK(k-1, loc\_indices, locations)$}\Comment{retrieve $k-1$ partition locations}
\State $Pattern \gets [\text{M copies of }selected[0]]$\Comment{initialize pattern to first selected level}
\For{$j \gets [0,...,k-1)$}\Comment{assign desired levels to rest of pattern}
\State $Pattern\left[splits[j]:splits[j+1]\right] \gets selected[j+1]$
\EndFor
\State Do something with the generated $Pattern$ (e.g. evaluate risk)
\EndFor
\EndFor
\EndFor
\EndProcedure
\end{algorithmic}
\end{algorithm}

\begin{algorithm}
\caption{Retrieve non-redundant sub-list from a list}\label{alg:nchoosek}
\begin{algorithmic}[1]
\Function{nchoosek}{$k,indices,optionList$}\Comment{retrieve $k$ elements from $optionList$ using $indices$ and update indices for next retrieval}
\State $ret\gets [] $
\For{$i\gets 0,...,k-1$} \Comment{Retrieve elements from $optionList$}
\State $ret.append(optionList[indices[i]]$)
\EndFor
\For{$i\gets k-1,...,0$} \Comment{update indices for next retrieval}
\If{$indices[i] < optionList.length-k+i$}
\State $indices[i] \gets indices[i] + 1$
\State \textbf{break}
\Else
\State mark $indices[i]$ as undetermined
\EndIf
\EndFor
\For{$i\gets 0,...,k-1$}
\If{$indices[i]$ is undetermined}
\State $indices[i] \gets indices[i-1]+1$ if $i > 0$ else 0
\EndIf
\EndFor
\State \textbf{return} $ret$
\EndFunction
\end{algorithmic}
\end{algorithm}

\clearpage

\bibliographystyle{IEEEtran}
\bibliography{references}